\documentclass[11pt, draftclsnofoot, onecolumn]{IEEEtran}
\usepackage{amsfonts}
\usepackage{amsmath,amssymb}
\usepackage{empheq}
\usepackage{epsfig,graphics,xcolor}
\usepackage{CJK}
\usepackage{setspace}
\usepackage{multirow}
\usepackage{cite}
\usepackage{algorithm, algorithmic}
\usepackage{array}
\usepackage{psfrag}
\usepackage{cases}
\usepackage[normalem]{ulem}
\ifCLASSOPTIONcompsoc
 \usepackage[tight, normalsize, sf, SF]{subfigure}
\else
 \usepackage[tight, footnotesize]{subfigure}
\graphicspath{{./Figures/}}

\newtheorem{proposition}{Proposition}
\newtheorem{assumption}{Assumption}
\newtheorem{definition}{Definition}
\newtheorem{remark}{Remark}

\newcommand{\st}[1]{{\color{green} \ifmmode\text{\sout{\ensuremath{#1}}}\else\sout{#1}\fi}}

\title{QoE Driven VR 360$^{\circ}$ Video Massive MIMO Transmission}

\author{Long~Teng, Guangtao~Zhai,~\IEEEmembership{Senior Member,~IEEE,} Yongpeng~Wu,~\IEEEmembership{Senior Member,~IEEE,} Xiongkuo~Min,~\IEEEmembership{Member,~IEEE}, Wenjun~Zhang,~\IEEEmembership{Fellow,~IEEE,} Zhi~Ding,~\IEEEmembership{Fellow,~IEEE,} and Chengshan~Xiao,~\IEEEmembership{Fellow,~IEEE}
\thanks{L. Teng, G. Zhai, Y. Wu, X. Min, and W. Zhang are with the Institute of Image Communication and Information Processing, Shanghai Jiao Tong University, Shanghai 200240, China (e-mail: tenglong@sjtu.edu.en; zhaiguangtao@sjtu.edu.cn; yongpeng.wu@sjtu.edu.cn; minxiongkuo@sjtu.edu.cn; zhangwenjun@sjtu.edu.cn). (\emph{Corresponding authors: G. Zhai \textrm{and} Y. Wu.}) }
\thanks{Z. Ding is with the Department of Electrical and Computer Engineering, University of California at Davis, Davis, CA 95616 USA (e-mail: zding@ucdavis.edu).}
\thanks{C. Xiao is with the Department of Electrical and Computer Engineering,
Lehigh University, Bethlehem, PA 18015 USA (e-mail: xiaoc@lehigh.edu).}
}

\begin{document}

\maketitle

\begin{abstract}
Massive  multiple-input and multiple-output (MIMO) enables ultra-high throughput
and low latency for tile-based adaptive virtual reality (VR) 360$^{\circ}$ video transmission in wireless network.
In this paper, we consider a massive MIMO system
where multiple users in a single-cell theater watch an identical VR 360$^{\circ}$ video.
Based on tile prediction, base station (BS) deliveries the tiles in predicted field of view (FoV) to users.
By introducing practical supplementary transmission
for missing tiles and unacceptable VR sickness,
we propose the first stable transmission scheme for VR video.
we formulate  an integer non-linear programming (INLP) problem to maximize
users' average quality of experience (QoE) score.
Moreover, we derive the achievable spectral efficiency (SE)
expression of predictive tile groups and the approximately
achievable SE expression of missing tile groups, respectively. Analytically, the overall throughput is related to the number of tile groups and the length of pilot sequences.
By exploiting the relationship between the structure of viewport tiles and SE expression, we propose a multi-lattice multi-stream grouping
method aimed at improving the overall throughput for VR video transmission.
Moreover, we analyze the relationship between QoE objective and number of predictive tile.
We transform the original INLP problem into an integer linear programming problem by setting the predictive tiles groups as some constants.
With variable relaxation and recovery, we obtain the optimal average QoE.
Extensive simulation results validate that the proposed algorithm effectively improves
QoE.
\end{abstract}

\begin{IEEEkeywords}
Tile-based adaptive VR 360$^{\circ}$ video, field of view (FoV), tolerant latency, tile grouping, massive MIMO, quality of experience (QoE), linear programming
\end{IEEEkeywords}
\IEEEpeerreviewmaketitle

\section{Introduction}
The recent explosive growth of smart devices and multimedia services
strongly motivate the development of
new technology to deliver virtual reality (VR) 360$^{\circ}$ video
across wireless networks.
The ultra-high resolution, high representation, panoramic scene, and multi-stimuli of VR provide a unique immersive experience,
allowing users to interact within an alternative world.
Unlike traditional video, panoramic scene of VR 360$^{\circ}$ video is captured by omnidirectional cameras. While watching VR videos,
users may freely adjust
orientation to retrieve expected immersive scene as part of
VR user interaction with support from interactive sensors \cite{HDM}.

As discussed in \cite{mag_VR}, transmitting VR 360$^{\circ}$ video, characterized by ultra-high data rate and low latency, presents critical challenges to wireless networking. Recent works have focused on
compressing the required data payload
in the area of VR 360$^{\circ}$ video processing. For example, VR 360$^{\circ}$ video is projected into a specific 2-D plane with multiple slices, and encoded with a established rule \cite{VR_projection}.
Generally, delivering all slices is unnecessary considering that the field of view (FoV) is limited. Transmitting only desired slices is able to decrease the data size to effectively relieve network load \cite{predicting_head}.
Despite such efforts,
transmitting ultra-high resolution VR video in real-time remains unrealistic
under limited wireless bandwidth and throughput.
Long transmission latency can cause human VR sickness.
An alternative proposal is to apply
content buffering or caching \cite{3C}, \cite{3C_fog} in wireless edge or device in advance. To this end,
content prediction is necessary. Buffering predicted contents in devices
before playback and adjusting the sequence of encoded segments to reduce response latency are likely to lessen the impact of random head movement \cite{optimal_viewport} and the VR sickness caused by stall \cite{mag_VR}.

Content prediction techniques in VR mainly include saliency prediction and quality assessment. Saliency \cite{bottom_up} can describe the importance of different visual contents and can be used to obtain the scope  of the most visually appealing areas automatically.
Quality assessment has also been widely used to analyze
VR 360$^{\circ}$ video content. For example,
\cite{MC360} proposes a blind image quality assessment model based on multi-channel convolutional neural network (CNN) architectures to
accumulate the objective quality scores of the VR 360$^{\circ}$ video, which can help derive the
probable slices of interest. The success of CNN in
slice prediction for VR 360$^{\circ}$ image \cite{Prediction_head}
and VR 360$^{\circ}$ video \cite{predicting_head}, \cite{VR_adaptation}
confirms the capability of learning based approach
according to the user behaviors in predicting the
``exact scope''. {In fact, we leverage the result of exact scope in next model and formulation.} In particular,
\cite{Gaze_aware} utilizes gaze-aware streaming
to limit the provisioning of high video quality to areas
near users' fixations, without quality loss in user perception.

Omnidirectional video coding is also an important element in VR.
High efficiency video coding (HEVC) \cite{HEVC_2012} is standardized collaboratively by a joint video exploration team (JVET) of ITU-T VCEG and ISO/IEC MPEG organizations, and the joint exploration model beyond HEVC developed by JVET provides a well-performing encoder with manageable complexity, with potential to improve the coding efficiency significantly.
In terms of sphere-to-plane coding on head mounted display (HMD), the work in \cite{Framework_coding} shows that equirectangular format saves 8.3$\%$ bit rate traffic. Considering the equirectangular format of VR 360$^{\circ}$ video, tile-based projection (TBP)
\cite{VR_adaptation}, \cite{fengjie,JET,VR_Modeling_QoE}, which splits the high resolution video into several tiles,
effectively reduces the transmitted data with low distortion
according to the viewport of user \cite{VR_projection}.
TBP is widely used in the projection process of VR video
and exhibits strong advantages in multicast application \cite{optimal_VR}\footnote{Compared with unicast, multicast can substantially improve the overall achievable throughput \cite{Co_pilot}.}.
Hence, in this work we apply the TBP with reasonable tile size
to transmit VR 360$^{\circ}$ video in wireless network.

{Most existing VR 360$^{\circ}$ video transmission schemes explore the optimization algorithms in a certain wireless network.
With optimal transmission time and power allocation, \cite{multicast_l} searches the multicast opportunity to respectively minimize the average transmission energy for the given video quality and maximize the video quality for the given energy budget in a time division multiple access (TDMA) system.
Based on the concept in \cite{multicast_l}, work in \cite{multi_multicast} exploits user transcoding and transcode-playback mode aimed to maximize the multicast opportunity with the consideration of smoothness requirement in a TDMA system.
\cite{optimal_pro} defines a performance metric for perfect, imperfect and unknown FoV probability distributions, and maximize the performance metric in a multi-carrier system.}

In VR video transmission, the quality of experience (QoE) is paramount.
Factors restricting QoE of VR video have been extensively
investigated in \cite{optimal_viewport}, \cite{VR_Modeling_QoE}, \cite{QoE_evaluation,VRSA,QoE_video,weight_2}.
Most previous works adopt mean opinion score based on
subjective quality evaluation.
Summarizing the works of \cite{optimal_viewport}, \cite{VR_Modeling_QoE}, \cite{QoE_evaluation,VRSA,QoE_video,weight_2}, there are three major factors
of VR quality that should be investigated
in wireless applications. The first is the
 overall tile quality perceived by multiple users,
which relates to the network capacity. The second is
the uncomfortable visual
perception caused by quality differences among tiles. The third
is the stall time caused by low transmission rate or data
retransmission. {Thus, we consider these three major factors in the QoE model.} Note that the stall time longer than
tolerant latency is a major cause for VR sickness.

To the best of our knowledge, the performance of existing wireless
transmission methods of VR 360$^{\circ}$ video has been less than satisfactory.
The major obstacle is the poor tile quality caused by the low link throughput. {Moreover, current works focusing on VR video transmission in wireless networks \cite{VR_adaptation},\cite{optimal_VR},\cite{QoE_evaluation},\cite{Taming_latency} generally assume that the exact FoV can be predicted infallibly from
machine learning and only transmit the predictied FoV}. {In addition, existing works, e.g., \cite{multicast_l,multi_multicast,optimal_pro} try to search the exact FoV through the viewing probability distribution, which leads to inaccurate results or requires many more tiles for transmission.} In practice, for the
reason of exceptional head movement caused by multi-stimuli \cite{VRSA},
e.g., when user is watching a scene with multi-stimuli like racing and
roller coaster, the FoV is difficult to be
predicted reliably and a much larger scope of tiles must be considered.
{In short, existing VR transmission works have not systematically considered the real supplementary transmission for missing tiles, which
is likely to cause unacceptable latency, and perceptual difference due to spatial quality variance. These
shortcomings present challenges to user QoE of VR delivered over wireless network.}

Massive multiple-input and multiple-output (MIMO) \cite{massive_MIMO},
can overcome effects of uncorrelated noise and fast fading
and deliver multiple streams to their respective users
simultaneously. Base stations (BSs) equipped with large-scale antenna arrays
can effectively exploit the estimated channel matrix \cite{JSDM}
to provide high sum-rate \cite{BDMA} and signal quality
\cite{MIMO_2.0}. Thus, massive MIMO has the potential to wirelessly
achieve the VR need for high-throughput and {low access latency \cite{MIMO_coding}}.
Moreover, the multiple tiles, treated as multiple streams, can be
easily transmitted to users simultaneously by taking advantage of the massive MIMO systems. Integrating massive MIMO within VR video transmission has strong potential to improve the QoE. Surprisingly, there has been
very few existing efforts in this direction.

{Existing works in \cite{multicast_l,multi_multicast,optimal_pro} have proven that grouping and multicast can efficiently improve network throughput. However, the multicast \cite{multicast_l,multi_multicast,optimal_pro} is uni-stream multicast in TDMA. Moreover, there has been no prior work that systematically combines the multi-stream multicast massive MIMO and VR 360$^{\circ}$ video transmission in the QoE optimization}. Motivated by the need for supplementary transmission for missing tiles
and the potential offered by massive MIMO, we consider a practical and innovative scenario involving QoE driven transmission of VR 360$^{\circ}$ video in multi-user massive MIMO wireless networks.
In this multicast setting, multiple users in a single-cell massive MIMO systems are engaged in the same VR 360$^{\circ}$ video.
Our goal is to implement the tile grouping and determine
the quantity of predictive tiles for maximizing the average QoE.
Specifically, the main contributions of this paper
are summarized as follows:
\begin{itemize}
  \item {We investigate the QoE driven VR 360$^{\circ}$ video transmission in multi-user massive MIMO systems, and systematically combine multi-stream multicast massive MIMO and VR 360$^{\circ}$ video transmission in the QoE optimization}. According to the real supplementary transmission for missing tiles and the unacceptable VR sickness,
 we propose a practical and stable transmission scheme. We formulate the average QoE objective of watching an identical VR 360$^{\circ}$ video.
  \item We derive a closed-form expression of the achievable spectral efficiency (SE) of predictive tile groups and the approximately achievable SE of missing tile groups under the maximum ratio transmission (MRT) and zero-forcing (ZF) precoding schemes, and allocate precoding power to guarantee consistent delivery rate of each stream based on max-min fairness (MMF).
  \item We analyze the relationship between SE and viewport tiles, and prove the existence of an optimal multi-stream grouping based on rectangular viewport. We further propose a multi-lattice multi-stream grouping (MLMSG) method to reduce the transmitted groups and pilot sequences during multicast.

  \item We adopt a variable number of predictive tiles. By setting the number of predictive tile groups as some constants, the original integer non-linear programming (INLP) problem is transformed into an integer linear programming (ILP) problem
  to optimize the final average QoE through relaxation and recovery. Extensive simulations demonstrate that the proposed algorithm
 effectively improves VR 360$^{\circ}$ video QoE at low complexity.
\end{itemize}

The remainder of this paper is organized as follows. We present the system model and problem formulation in Section \ref{system_model}. Section \ref{SE_F} derives the achievable SE of each group tile in massive MIMO system. And we propose the MLMSG in Section \ref{tile_grouping}. In Section \ref{QoE}, we maximize the average QoE by turning the non-linear problem into a linear problem. Simulation results are presented in Section \ref{simulation} to evaluate the performance of our proposed algorithm. We finally conclude our paper in Section \ref{conclusion}.

\textit{Notations}: Lower case, boldface lower case, and boldface upper case  letters denote scalars, vectors, and matrices, respectively; $\mathbf{I}_N$ denotes the identity matrix of size $N$. $\mathbf{x} \thicksim \mathcal{CN}(\textbf{0},\Sigma)$ indicates that $\mathbf{x}$ is a circularly symmetric complex Gaussian vector with zero mean and covariance matrix $\Sigma$.
The superscripts $(\cdot)^T$, $(\cdot)^{*}$, and $(\cdot)^H$ stand for the transpose, conjugate, and  conjugate-transpose of a matrix, respectively. We use $\mathbb{E}\{\cdot\}$ to denote ensemble expectation and $|{\mathbf{x}}|$ to represent cardinality of a set ${\mathbf{x}}$. $\lceil x \rceil$ and $\lfloor x \rfloor$ stand for the smallest integer larger than or equals to $x$ and the largest integer smaller than or equals to $x$, respectively.
\section{System Model And Problem Formulation}\label{system_model}
In this section, we introduce the system model which contains the tile-based adaptive regime in VR video processing and the considered deployment scenario in wireless network. Then the problem formulation maximizing the average QoE is presented.
\subsection{Tile-Based Adaptive Regime}
\begin{figure}[htb]
\centering
\subfigure[Head rotation direction]{\label{HMD}
    \includegraphics[width=0.3\textwidth]{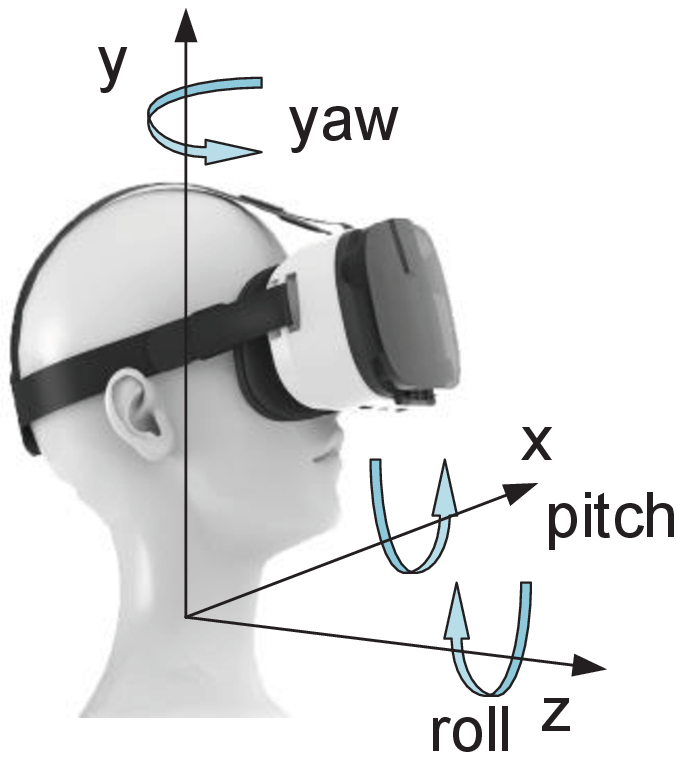}
      }
     \hspace{0.1in}
\subfigure[Equirectangular format and FoV format]{\label{tiling}
    \includegraphics[width=0.3\textwidth]{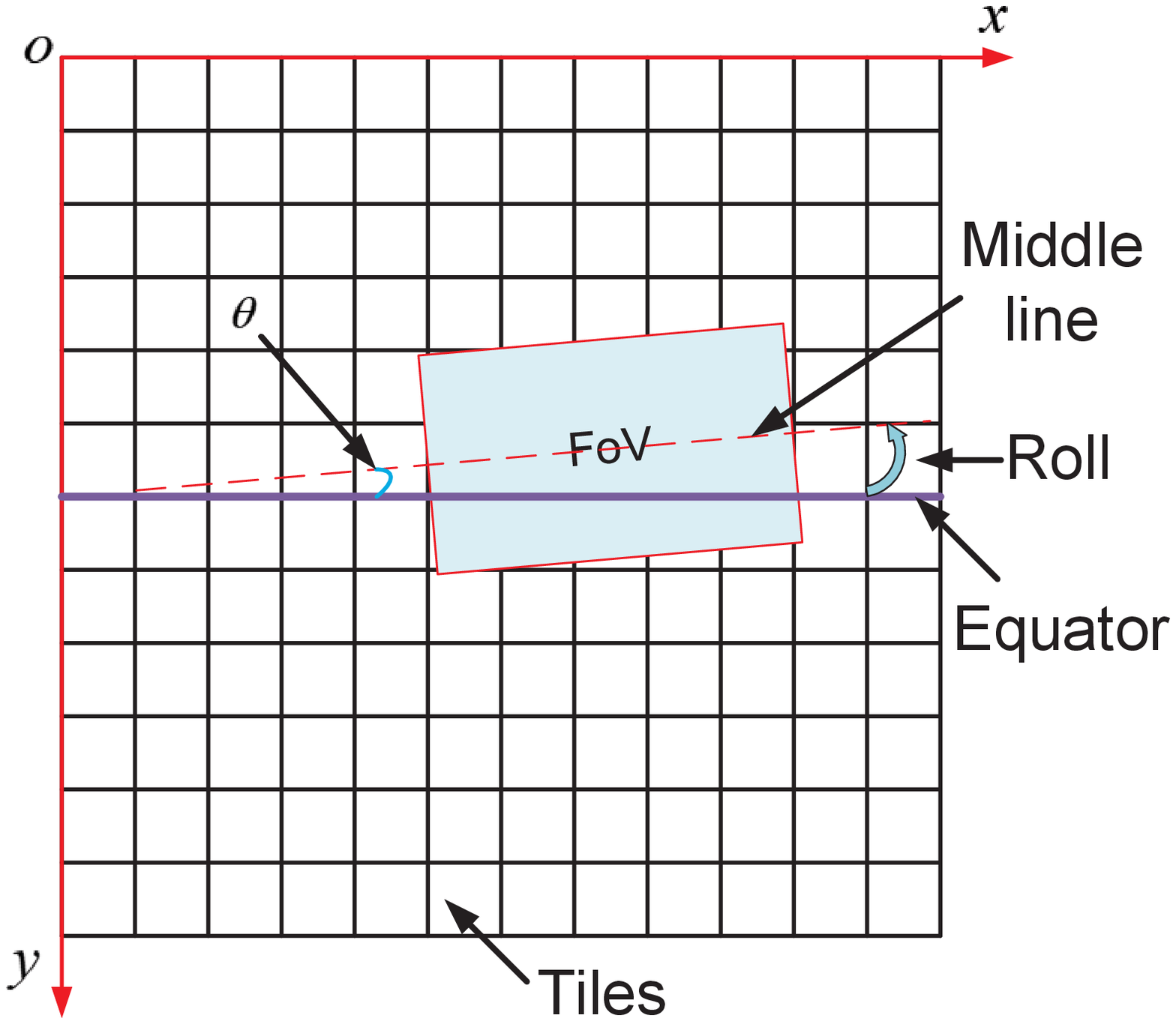}}
\caption{(a) Head rotation model; and (b) equirectangular projection based on tiling approach.}
\end{figure}

VR 360$^{\circ}$ video captured and stitched by omnidirectional camera has a spherical shape in the original format. For a watching user, the scene within viewport is displayed in the HMD, and user can turn their head and eyes to track the interesting contents as illustrated in Fig. \ref{HMD}. Utilizing the typical tilling approach \cite{VR_projection}, the whole spherical streaming is projected into an equirectangular format with multiple sized tiles as shown in Fig. \ref{tiling}, which can be encoded according to a set of quality levels. It is noted that the equator of VR sphere is projected into the horizontally intermediate line of equirectangular. Denote the coordinate origin as $O$, the horizontal coordinate set as $\mathcal{H}$, and the vertical coordinate set as $\mathcal{V}$, then the tile index in the equirectangular can be represented by ${\zeta(x,y)}, x\in\mathcal{H},y\in \mathcal{V}$. Without moving head, FoV covering 150$^{\circ}$ horizontally and 120$^{\circ}$ vertically including eyes movement is recommended in \cite{mag_VR}. Accordingly, the tiles contained FoV are encoded in high quality and the remaining ones can be encoded in basic low-quality or abandoned to improve the QoE under limited network capacity. Simultaneously, HMD sensors can feel user movement and activities, and can provide helpful information to predict the desired tiles and implement tile-based adaption.
\subsection{Deployment Scenario}
\begin{figure*}[htb]
\centering
    \includegraphics[width=0.35\textwidth]{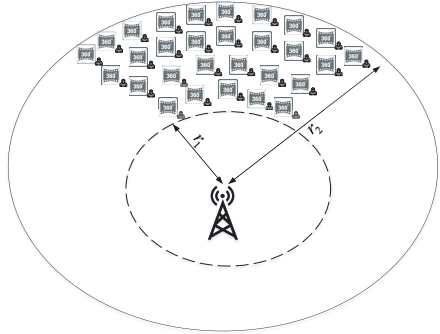}
      \caption{Multiple users experiencing VR 360$^{\circ}$ video in a massive MIMO cell.}\label{scenario}
\end{figure*}

{We consider an open VR theater with seats arranged in multiple cycles in a single-cell, where each user wears a single-antenna HMD and seats on a rotatable but fixed chair.} There are $K$
active users indexed by set $\mathcal{K}=\{1,\;\cdots,\;K\}$ {and the transmission bandwidth is $W$}. The scenario is illustrated in Fig. \ref{scenario}, where the BS equipped with $N$ antennas locates in the center to serve the $K$ active users simultaneously in the time division duplexing (TDD) mode. The powers of HMD and BS are denoted as $P_{{u}}$ and $P_{{d}}$, respectively. In the scenario, the radii of inner cycle and outer cycle are $r_1$ and $r_2$, respectively.

Assume a block-fading channel model
which remains invariant in each coherence interval $T$, where $T$ is the product of the coherence bandwidth $C_B$ and coherence time $C_T$. {Further, the duration of exceptional head movement is relative small compared with the coherence time such that Doppler frequency offsets can be negligible.} In the system, we consider uncorrelated Rayleigh fading channel responses, and denote $\mathbf{h}_k$ as the channel response of user $k$, i.e., $\mathbf{h}_k\sim \mathcal{CN}(\textbf{0},{\psi}_k \mathbf{I}_N)$, where ${\psi}_k$ is the large-scale fading coefficient. Note that practical channels might have spatially correlated fading or line of sight components, but theoretical studies and practical measurements carried out in real massive MIMO propagation environments have shown that SE can be predicted using uncorrelated fading models \cite{UM}. Moreover, this channel model enables us to present novel insights into VR 360$^{\circ}$ video massive MIMO transmission.

Within each coherence interval, we focus on uplink pilot transmission and downlink data transmission. During uplink pilot transmission, users send uplink pilots to enable BS to estimate their respective uplink channels. The pilots in classic unicast massive MIMO system are orthogonal. Applying the concept of co-pilot proposed in \cite{Co_pilot}, the users assigned to receive the same tile would share a pilot in each multicast stream. It is therefore reasonable to assume that the pilots of different streams are orthogonal. Taking advantage of reciprocity between uplink and downlink channels in TDD, the BS performs downlink precoding based on the estimated channels and deliveries the tile.

\begin{figure*}[htb]
\centering
    \includegraphics[width=0.3\textwidth]{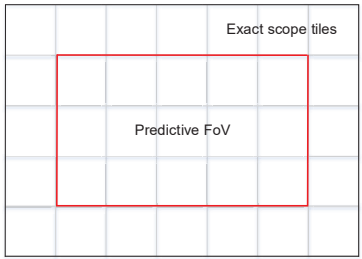}
      \caption{Representations of predictive FoV and exact scope.}\label{scope}
\end{figure*}

The FoV can move arbitrarily by rotation in the directions of pitch, yaw, and roll. Denote the angle between the middle line of viewport and equator as $\theta$, which is influenced by the rotation in the direction of roll.
We illustrate the relationship between $\theta$ and roll direction in Fig. \ref{tiling}. It was shown in \cite{VR_saliency} that a large proportion of fixation distributes near the
equator, and the authors of \cite{VR_adaptation} stated that rotation in the direction of roll is negligible compared with the other two directions. Hence, we mainly focus on $\theta=0$ where predictive FoV tiles are rectangle in shape.
Further, the number of tiles different
between the FoV and the exact scope is directly and positively correlated to the distribution area of multi-stimuli around a certain viewport \cite{predicting_head}. {Over 80$\%$ prediction accuracy can be obtained by machining learning \cite{predicting_head},\cite{Flare}. Thus the exact scope that can be predicted is consequently a little larger than the FoV.} In addition, the structural similarity proposed in \cite{SSIM} recommends that the predictive tiles
reside in the middle of the exact scope. We apply this concept
in tile buffering. For clarity, we illustrate the case in Fig. \ref{scope}. Without fully accurate prediction, BS transmits desired missing tiles, i.e., the tiles within exact scope
outside the predictive set, to supplementally
meet user needs. The
 BS has completely cached the original VR 360$^{\circ}$ video and the HMD has ability of buffering and computing.
The BS leverages the existing CNN prediction model to calculate the prospective viewports of each user, and transmits the corresponding data to HMDs in advance.
Having received the transmitted data, the HMD selectively arranges the tiles, stitches 2-dimensional (2D) tiles into 3-dimensional (3D) FoV, renders and displays the expected scenes.

Like traditional video, VR 360$^{\circ}$ video has similar frame structure and frame size. Based on characteristics of tile prediction, time interval of predictive tiles\footnote{Note
that the time interval of predictive tiles is also the
	corresponding playback time.} should be scheduled
reasonably to avoid
 non-real-time transmission or excessive overhead
on computing time and energy.
Further, time interval of missing tiles should be short to reduce the stall time. In this work we set $T_{1}$ and $T_{2}$ as the time intervals of predictive tiles and missing tiles, respectively. All tiles in each interval time include three parts, i.e., the predictive tiles within $T_1$, denoted as $Z_0$, the missing tiles within $T_2$, denoted as $Z_1$, and the subsequent missing tiles within $T_1-T_2$, denoted as $Z_2$.
Note the transmission order: $Z_0$, $Z_1$, and $Z_2$. Without being content specific, quality level of each
tile strictly relates to the set of encoding rates. We
denote the encoding rates of predictive tile and missing
tile as ${\eta}_p$, and ${\eta}_m$, respectively, which
belong to encoding rate set $\mathcal{R}=\{R_1,\cdots,R_d,\cdots,R_D\}$. For public VR theater, VR sickness due to stalling is unacceptable.
Thus, there is
an upper bound of tolerable stall time $T_{y}$. In addition, assuming user fairness, we make the following assumptions.
\begin{assumption}\label{assumption}
\begin{enumerate}
\item The encoding rates of predictive tiles and missing tiles for every user are the same;
\item The numbers of predictively transmitted tiles of
each frame are the same for each user,
denoted as $N_p^k=N_p, \forall k\in \mathcal{K}$;
thus the expected numbers of missing tiles of each frame
are also the same for each user, denoted by $N_m^k=N_m,\forall k\in \mathcal{K}$.
\end{enumerate}
\end{assumption}

\subsection{Problem Formulation}

In this paper, we maximize the average QoE of VR 360$^{\circ}$ video transmission in massive MIMO systems by joint consideration among $N_p$, $N_m$, ${\eta}_p$, and ${\eta}_m$.

\begin{figure*}[htb]
\centering
    \includegraphics[width=0.6\textwidth]{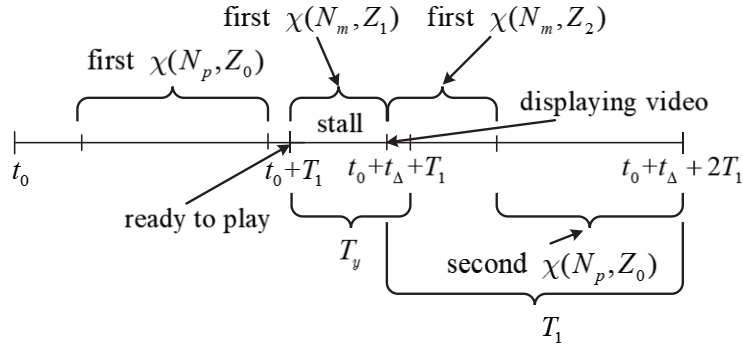}
      \caption{Stable transmission scheme for $Z_0$, $Z_1$, and $Z_2$.}\label{interval}
\end{figure*}

{Denote functions $\chi(N_p,Z_0)$, $\chi(N_m,Z_1)$, and $\chi(N_m,Z_2)$ as the transmission latencies of transmitting $Z_0$, $Z_1$, and $Z_2$, respectively}, where the size of $Z_0$, $Z_1$, and $Z_2$ are $T_1\eta_p$, $T_2\eta_m$, and $(T_1-T_2)\eta_m$, respectively. To avoid VR sickness caused by stall,
\begin{align}
\chi(N_m,Z_1)\leq T_y\label{time_sick}.
\end{align}
In addition, during the current playback time $T_1$, the transmission of last $Z_2$ will occupy the current transmission time for current $Z_0$. For clarity, we illustrate the scheme in Fig. \ref{interval}.
The first $Z_2$ and the second $Z_0$ should be delivered within $T_1$, to maintain transmission model stability. Under the identical prediction model and smooth short interval $T_1$, the number of missing tiles between two adjacent time intervals are approximately the same.
For this reason, we can reasonably assume that
$Z_0$ and $Z_2$ in the same interval time should be
delivered within $T_1$. In other words, we require
\begin{align}
\chi(N_p,Z_0) + \chi(N_m,Z_2)\leq T_1.\label{time_all}
\end{align}
The prediction error is relatively small based on the
existing prediction model. Hence, the number of missing
tiles is smaller than the number of tiles hit by prediction.
 Based on the visual perception and joint consideration
 between $\chi(N_p,Z_0)$ and $\chi(N_m,Z_1)$,
 we set
\begin{align}
\eta_m\leq \eta_p \label{pre_com}.
\end{align}
{The QoE in \cite{weight_2} is formulated by the weighted average of video quality minus the weighted spatial video quality. Generally, the distortion of each tile depends on the encoding rate while the spatial video quality can be determined by $N_m\cdot(\eta_p-\eta_m)$.
As the number of hit tiles is greater than that of missing tiles, the encoding rate of predictive tiles is the quality of major tiles, which is near the average video quality.
Thus, the average mean squared error and the spatial quality variance can be controlled by $\eta_p$ and $N_m\cdot(\eta_p-\eta_m)$ jointly through weight adjustment.
A QoE score is intuitively formulated
as the encoding rate of predictive tiles subtracting the perceptual difference, a penalty factor.}
Based on Assumption \ref{assumption}, we can assess QoE in a video frame within $T_1$.
Hence, the optimization objective is formulated as
\begin{align}
\left(\textbf{P0}\right) \quad \max_{\eta_p,\eta_m}\quad & \frac{1}{K}\sum_{k=1}^{K}\alpha_k \eta_p-\beta_k N_m
\cdot (\eta_p-\eta_m) \nonumber \\
\textrm{s.t.}\quad & \eqref{time_sick},\eqref{time_all},\eqref{pre_com} \nonumber\\
& \eta_m, \eta_p \in \{R_1,\cdots,R_d,\cdots,R_D\}\label{RD}
\end{align}
where for user $k$, we assign weights
$\alpha_k$ and $\beta_k$
for the encoding rate of predictive tiles and
the encoding rate difference $(\eta_p-\eta_m)$,
respectively. {Particularly, based on human visual system and saliency influence in \cite{Prediction_head,VR_saliency}, larger saliency degree in VR FoV carries out larger visual impact, especially in the case of perceptual difference. Thus, $\alpha_k$ and $\beta_k$ are positively related to saliency degree, especially for $\beta_k$.}

\emph{Problem analysis}: In the multi-user massive MIMO systems,
variables $N_p$, $N_m$, and transmission mode jointly
determine
$\chi(N_p,Z_0)$, $\chi(N_m,Z_1)$, and $\chi(N_m,Z_2)$, which further decide the average QoE score. The existing basic transmission mode, {e.g., uni-stream multicast in \cite{multicast_l},} which counts
indices of tiles for all active users
and transmits each tile by uni-stream multicast,
is unable to meet the needs of VR 360$^{\circ}$ video transmission.
Thus, optimizing the
transmission mode for VR 360$^{\circ}$ video in massive MIMO systems is necessary. Normally, optimized
transmission mode in massive MIMO systems is related
to the tile grouping method of multiple tiles, which
aims to maximize the overall network throughput.

Under a certain transmission mode, integer $\eta_m$ in constraint \eqref{time_sick} is linear with integer $N_m$,
whereas integers $\eta_p$ and $\eta_m$ in constraint
\eqref{time_all} are linear with respect to integers $N_p$ and $N_m$. Furthermore, $N_m(\eta_p-\eta_m)$ is a product
of integers $N_m$ and $\eta_p-\eta_m$.
Hence, the problem (\textbf{P0}) is a high complexity
INLP problem.
Solving this INLP problem directly would
be impractical for real-time transmission.
Hence, we consider low complexity alternatives
by jointly considering
transmission mode, variables $N_p$ and $N_m$.

To solve problem (\textbf{P0}) efficiently and approximately,
we first derive the achievable spectral efficiency to establish
a basic foundation to optimize tile grouping.
We then examine the relationship between $N_p$ and $N_m$,
as well as the product $N_m(\eta_p-\eta_m)$
to efficiently optimize the QoE.

\section{Achievable Spectral Efficiencies in Tile Transmission}\label{SE_F}
In this section, we analyze the SE of VR video {by the major linear MRT and ZF precoding} in massive MIMO systems. In massive MIMO systems, the tile transmission is either unicast or multicast, and the transmitted data are either uni-stream or multi-stream. We focus on the multi-stream multicast in the VR video transmission process.
We index the multiple groups
formed by all transmitted tiles of each frame by
the index set $\mathcal{G}=\{1,\cdots,g,\cdots,G\}$, and the multiple streams in each group queue for transmission.
{We indicate the indices of users requesting tile in group $g$ by a signal matrix $\mathbb{J}_g$ with $F$ rows and $B$ columns, where the $(f,b)$-th
entry $\mathbb{J}_g(f,b)$ represents the index of the $b$-th user that requests the $f$-th tile stream in group $g$.
We further use ${\mathbb{j}_{g,f}}, f=1,\cdots,F$, to represent the set of users in the $f$-th row of $\mathbb{J}_g$.}

\subsection{Channel Estimation}
We define a pilot matrix $\mathbf{\Phi}_{g}=\sqrt{\sigma_g}\left[\boldsymbol{\phi}_{g,1}, \cdots,\boldsymbol{\phi}_{g,f},\cdots,\boldsymbol{\phi}_{g,F}\right]$, composed by $\sigma_g$ mutually orthogonal $\sigma_g$-length pilot sequences, where $\boldsymbol{\phi}_{g,f}$ is a pilot sequence for {each user in $\mathbb{j}_{g,f}$}. The received uplink signal at the BS is
\begin{align}\label{estimation}
\mathbf{Y}_g=\mathbf{H}_g\mathbf{\Phi}_g+\mathbf{N}_g
\end{align}
where matrix $\mathbf{H}_g$ is formed by
entries $\mathbf{H}_g(f,b)$. Each entry
$\mathbf{H}_g(f,b)$ represents the channel response of user $\mathbb{J}_g(f,b)$, and
$\mathbf{N}_g\in \mathbb{C}^{N\times \sigma_g}$ is the normalized additive noise matrix with
entries  $\mathbf{N}_g (t,s) \thicksim \mathcal{CN}(0,1)$. Thus, we have
\begin{align}\label{estimation2}
\mathbf{Y}_g=\sum_{i=1}^{F}\sum_{b={1}}^{B}
\sqrt{\sigma_gq^u}
\mathbf{H}_{g}(i,b)\boldsymbol{\phi}^T_{g,i}+\mathbf{N}_g
\end{align}
where $q^u$ is the normalized uplink power. We can
obtain the received sequence from $\mathbf{Y}_{g}$ via
\begin{align}
\mathbf{y}_{g,f}= \mathbf{Y}_g \boldsymbol{\phi}_{g,f}^*=\sum_{b=1}^{B}\sqrt{\sigma_gq^u}\mathbf{H}_g(f,b) +\mathbf{n}_{g,f}
\end{align}
where $\mathbf{n}_{g,f}\sim \mathcal{CN}(\textbf{0},\mathbf{I}_N)$ is normalized additive noise vector corresponding to
the users in {$\mathbb{j}_{g,f}$}. According to the MMSE estimation proposed in \cite{Kay:estimation}, BS can estimate the channel response ${\mathbf{H}}_g(f,b)$ as follows
\begin{align}\label{estimation_mu1}
\tilde{{\mathbf{H}}}_g(f,b)=\frac{\sqrt{\sigma_gq^u}{\boldsymbol{{\Psi}}}_g(f,b)}{1+\sum_{t=1}^{B}\sigma_g q^u{\boldsymbol{{\Psi}}}_g(f,t)}\left(\sum_{t=1}^{B}\sqrt{\sigma_gq^u}{\boldsymbol{{\Psi}}}_g(f,b) {\mathbf{H}}_g(f,t)+\mathbf{n}_{g,f}\right)
\end{align}
where {${\boldsymbol{{\Psi}}}_g(f,b)$ is the large-scale fading coefficient of $\mathbb{J}_g(f,b)$, and}
$\tilde{{\mathbf{H}}}_g(f,b)\sim \mathcal{CN}(\textbf{0},{\mathbf{U}}_g(f,b)\mathbf{I}_N)$ with ${\mathbf{U}}_g(f,b)=\frac{\sigma_gq^u\left(\boldsymbol{{\Psi}}_g(f,b)\right)^2}{1+\sum_{t=1}^{B}\sigma_g q^u{\boldsymbol{{\Psi}}}_g(f,t)}$. Due to the linear combination, we estimate ${\mathbf{h}}_{g,f}$, the channel response {for $\mathbb{j}_{g,f}$}, to be $\sum_{t=1}^{B}\sqrt{\sigma_gq^u}{\mathbf{H}}_g(f,b)$ and we have
\begin{align}\label{estimation_mu2}
\tilde{{\mathbf{h}}}_{g,f}=\frac{\sum_{t=1}^{B}\sigma_gq^u{\boldsymbol{{\Psi}}}_g(f,t)}{1+\sum_{t=1}^{B}\sigma_g q^u{\boldsymbol{{\Psi}}}_g(f,t)}\left(\sum_{t=1}^{B}\sqrt{\sigma_gq^u}{\boldsymbol{{\Psi}}}_g(f,b) {\mathbf{H}}_g(f,t)+\mathbf{n}_{g,f}\right)
\end{align}
where
$\tilde{{\mathbf{h}}}_{g,f}\sim \mathcal{CN}(\textbf{0},\mu_{g,f}\mathbf{I}_N)$ with $\mu_{g,f}=\frac{\left(\sum_{t=1}^{B}\sigma_gq^u{\boldsymbol{{\Psi}}}_g(f,b)\right)^2}{1+\sum_{t=1}^{B}\sigma_g q^u{\boldsymbol{{\Psi}}}_g(f,t)}$. Thus, we have
\begin{align}\label{estimation_mu3}
\tilde{{\mathbf{H}}}_g(f,b)=\frac{\sqrt{\sigma_gq^u}{\boldsymbol{{\Psi}}}_g(f,b)}{\sum_{t=1}^{B}\sigma_gq^u{\boldsymbol{{\Psi}}}_g(f,t)}\tilde{{\mathbf{h}}}_{g,f}.
\end{align}

Note that unicast has the same derivation, and the difference is that $\sum_{t=1}^{B}\sigma_gq^u{\boldsymbol{{\Psi}}}_g(f,t)$ in \eqref{estimation_mu1}-\eqref{estimation_mu3} is equal to $\sigma_gq^u{\boldsymbol{{\Psi}}}_g(f,b)$.
\subsection{Achievable Spectral Efficiency in Downlink Transmission}
The received sequence of users in group $g$ is
\begin{align}\label{downlink}
\mathbf{r}_{g}=\mathbf{H}^H_g\mathbf{B}_g\mathbf{{s}}_g+\mathbf{N}_g
\end{align}
where $\mathbf{s}_g=[{s}_{g,1},\cdots,{s}_{g,f},\cdots,{s}_{g,F}]^H$ represents the transmitted sequence of data symbols and $\mathbf{B}_g=[\mathbf{b}_{g,1},\cdots,\mathbf{b}_{g,F}]$ is the precoding matrix of group $g$ in the system. Hence, the received signal of user $\mathbb{J}_g(f,b)$ is
\begin{align}\label{receive}
{\textbf{r}_g(f,b)}={\mathbf{H}_g(f,b)^H}\mathbf{B}_g\mathbf{{s}}_g+{\mathbf{N}_g(f,b)}.
\end{align}
\subsubsection{MRT Precoding}
The precoding vector for the tile to $\mathbb{j}_{g,f}$ is
\begin{align}\label{power_MRT}
\mathbf{b}_{g,f}^{\textrm{MRT}}=\sqrt{\frac{q^d_{g,f}}{N\mu_{g,f}}}\tilde{\mathbf{h}}_{g,f}
\end{align}
where $q_{g,f}^d$ is the downlink power of the precoding vector for $\mathbb{j}_{g,f}$.
The received signal of user $\mathbb{J}_g(f,b)$ is
\begin{align}
{\textbf{r}^{\textrm{MRT}}_g(f,b)}={\mathbf{H}}_g(f,b)^H\mathbf{b}^{\textrm{MRT}}_{g,f}{s}_{g,f}+\sum_{i=1,i\neq f}^{F}{\mathbf{H}}_g(f,b)^H\mathbf{b}^{\textrm{MRT}}_{g,i}{s}_{g,i}+{\mathbf{N}_g(f,b)}.
\end{align}
Hence, the signal-to-interference-plus-noise-ratio (SINR) of user $\mathbb{J}_g(f,b)$ is
\begin{align}
{\boldsymbol{\Omega}^{\textrm{MRT}}_g(f,b)}=\frac{\left|\mathbb{E}\left\{{\mathbf{H}_g(f,b)^H}\mathbf{b}^{\textrm{MRT}}_{g,f}\right\}\right|^2} {1-\left|\mathbb{E}\left\{{\mathbf{H}_g(f,b)^H}\mathbf{b}^{\textrm{MRT}}_{g,f}\right\}\right|^2+\sum_{i=1}^{F} \mathbb{E}\left\{\left|{\mathbf{H}_g(i,b)^H}\mathbf{b}^{\textrm{MRT}}_{g,i}\right|^2\right\}}.
\end{align}
Based on the derivation in \cite{UM}, ${\boldsymbol{\Omega}^{\textrm{MRT}}_g(f,b)}$ is
\begin{align}\label{SINR}
{\boldsymbol{\Omega}^{\textrm{MRT}}_g(f,b)}=\frac{Nq_{g,f}^d{\mathbf{U}}_g(f,b)}{1+{\boldsymbol{{\Psi}}}_g(f,b)P}
\end{align}
where $P$ is the total normalized downlink power. And the corresponding SE is
\begin{align}\label{SE}
{\boldsymbol{\Gamma}^{\textrm{MRT}}_g}(f,b)=(1-\frac{\sigma_g}{T})\textrm{log}_2(1+{\boldsymbol{\Omega}^{\textrm{MRT}}_g(f,b)}).
\end{align}

\subsubsection{ZF Precoding}
The precoding vector for tile $g_f$ is
\begin{align}\label{power_ZF}
\mathbf{b}_{g,f}^{\textrm{ZF}}=\sqrt{(N-\sigma_g)q^d_{g,f}\mu_{g,f}}\tilde{\mathbf{H}}_g \left(\tilde{\mathbf{H}}_g^H\tilde{\mathbf{H}}_g\right)^{-1}\mathbf{e}_{g,f}
\end{align}
where $\mathbf{e}_{g,f}$ is the $f$-th column of a identity matrix $\mathbf{I}_{\sigma_g}$.
The received signal is
\begin{align}
\mathbf{r}_{g}^{\textrm{ZF}}= \sum_{i=1}^{F}{s}_{g,i}\mathbf{H}_{g}^H\mathbf{b}_{g,i}^{\textrm{ZF}}+\mathbf{N}_g.
\end{align}
By replacing $\mathbf{H}_{g}^H$ with $\mathbf{Z}\tilde{\mathbf{H}}_{g}^H-\hat{\mathbf{H}}_{g}^H$, where $\hat{\mathbf{H}}_{g}$ and $\mathbf{Z}$ are the estimation error $\mathbf{H}_g-\tilde{\mathbf{H}}_{g}$ and a diagonal matrix $\left[\frac{\sqrt{\sigma_gq^u}{\boldsymbol{{\Psi}}}_g(1,1)}{\sum_{i=1}^{B}\sigma_gq^u{\boldsymbol{{\Psi}}}_g(1,i)}, \cdots,\frac{\sqrt{\sigma_gq^u}{\boldsymbol{{\Psi}}}_g(F,B)}{\sum_{w=1}^{B}\sigma_gq^u{\boldsymbol{{\Psi}}}_g(F,w)}\right]$ based on \eqref{estimation_mu3}, respectively,
$\mathbf{r}_{g}^{\textrm{ZF}}$ turns to be
\begin{align}\label{Cerror}
\mathbf{r}_{g}^{\textrm{ZF}}=\sum_{i=1}^{F} {s}_{g,i} \left(\mathbf{Z}\tilde{\mathbf{H}}_{g}^H\mathbf{b}_{g,i}^{\textrm{ZF}}-\hat{\mathbf{H}}_{g}^H \mathbf{b}_{g,i}^{\textrm{ZF}}\right)+\mathbf{N}_g.
\end{align}
In \eqref{Cerror}, {matrix $\hat{\mathbf{H}}_g$ is formed by entries $\hat{\mathbf{H}}_g(f,b)$}, {where entry $\hat{{\mathbf{H}}}_g(f,b)$ is the estimation error of channel response for $\mathbb{J}_g(f,b)$.}
$\tilde{\mathbf{H}}_{g}^H\mathbf{b}_{g,i}^{\textrm{ZF}}$ is equal to $\sqrt{(N-\sigma_g)q^d_{g,f}\mu_{g,f}} \mathbf{e}_{g,i}$ according to \eqref{power_ZF}; thus, we have
\begin{align}
{\textbf{r}^{\textrm{ZF}}_g(f,b)}=\frac{\sqrt{(N-\sigma_g)q^d_{g,f}\mu_{g,f}\sigma_gq^u} {\boldsymbol{{\Psi}}}_g(f,b)}{\sum_{t=1}^{B}\sigma_gq^u{\boldsymbol{{\Psi}}}_g(f,t)}{s}_{g,f}-\sum_{i=1}^{F}\mathbf{\hat{\mathbf{H}}}_g(f,b)^H \mathbf{b}_{g,i}^{\textrm{ZF}}{s}_{g,i}+{\mathbf{N}_g(f,b)}.
\end{align}
Based on the derivation in \cite{UM}, {the SINR of user $\mathbb{J}^{\textrm{ZF}}_g(f,b)$} is
\begin{align}
{\boldsymbol{\Omega}^{\textrm{ZF}}_g(f,b)}=\frac{\left(N-\sigma_g\right)q_{g,f}^{{d}}{\mathbf{U}}_g(f,b)}{1+P({\boldsymbol{{\Psi}}}_g(f,b)-{\mathbf{U}}_g(f,b))}.
\end{align}
And the corresponding SE is
\begin{align}\label{SE_ZF}
{\boldsymbol{\Gamma}^{\textrm{ZF}}_g(f,b)}=(1-\frac{\sigma_g}{T})\textrm{log}_2(1+{\boldsymbol{\Omega}^{\textrm{ZF}}_g(f,b)}).
\end{align}

Note that uni-stream transmission has the same derivation, and the difference is that $\mathbf{B}_g$ and $\mathbf{{s}}_g$ in \eqref{downlink}  are a single vector and a one-dimensional data symbol, respectively.
\subsection{Max-Min Fairness}
In a multi-stream group, the common performance metric is MMF, where we want to maximize the minimal SINR among the streams. For MRT, the target is $\max_{q_{g,f}^d} \min_{{\mathbf{A}}_g^{\textrm{MRT}}(f,b)} {\mathbf{A}}_g^{\textrm{MRT}}(f,b) q_{g,f}^d$, where {${\mathbf{A}}_g^{\textrm{MRT}}(f,b)=(N{\mathbf{U}}_g(f,b))/(1+{\boldsymbol{{\Psi}}}_g(f,b)P)$, is the $(f,b)$-th entry of a power coefficient matrix $\textbf{A}_g$}.
For the {user set $\mathbb{j}_{g,f}$}, we extract the minimal value in the set $\left\{{\mathbf{A}}_g^{\textrm{MRT}}(f,1),\cdots,{\mathbf{A}}_g^{\textrm{MRT}}(f,b),{\mathbf{A}}_g^{\textrm{MRT}}(f,B)\right\}$ as the minimal power coefficient $a_{g,f}^{\textrm{MRT}}$, yielding the solution $q_{g,f}^{{d}}=\frac{P}{a_{g,f}^{\textrm{MRT}}}/\sum_{t=1}^{F}\frac{1}{a_{g,t}^{\textrm{MRT}}}$. Likewise, the power coefficient ${\mathbf{A}}_g^{\textrm{ZF}}(f,b)$ is
$\left(N-\sigma_g\right)q_{g,f}^{{d}}{\mathbf{U}}_g(f,b)/\left[1+({\boldsymbol{{\Psi}}}_g(f,b)-{\mathbf{U}}_g(f,b))P\right]$, and we also extract their minimal value as the minimal power coefficient $a_{g,f}^{\textrm{ZF}}$, yielding the solution  $q_{g,f}^{{d}}=\frac{P}{a_{g,f}^{\textrm{ZF}}}/\sum_{t=1}^{F}\frac{1}{a_{g,t}^{\textrm{ZF}}}$.

Hence, {the SINRs of each stream in group $g$ by either MRT precoding or ZF precoding are the same, which are denoted as $\Omega_{g,\epsilon}^{\textrm{MRT}}=\textrm{log}_2(1+P/\sum_{t=1}^{F}\frac{1}{a_{g,t}^{\textrm{MRT}}})$ and $\Omega_{g,\epsilon}^{\textrm{ZF}}=\textrm{log}_2(1+P/\sum_{t=1}^{F}\frac{1}{a_{g,t}^{\textrm{ZF}}})$, respectively.}
The achievable SEs of each stream in group $g$ are also the same, which are denoted as {${\Gamma}_{g,\epsilon}^{\textrm{MRT}}=(1-\sigma_g/T)\Omega_{g,\epsilon}^{\textrm{MRT}}$ and ${\Gamma}_{g,\epsilon}^{\textrm{ZF}}=(1-\sigma_g/T)\Omega_{g,\epsilon}^{\textrm{ZF}}$, respectively.} Note that the rule for predictive groups is also applicable to missing groups, and we denote the missing group set as $\mathcal{J}=[1,\cdots,j,\cdots,J]$.

\section{Multi-Stream Grouping Based on Viewport}\label{tile_grouping}
In this section, we try to search the optimal multi-stream grouping in VR video massive MIMO systems, the SE of which is derived in Section \ref{SE_F}. According to the equations \eqref{SE} and \eqref{SE_ZF}, we have that decreasing the group number $G$ and the length of pilot sequence $\sigma_g$ is the key to maximize the systems throughput. We analyze both the characteristics and constraints of multi-streaming group, and proposed a multi-stream grouping method based on multiple viewports. Note that the predictive tiles can be obtained; thus we firstly analyze the predictive tiles.

\subsection{Multi-Stream Group Based on Viewport Tiles}
According to the assumption that the number of predictive tiles among users is the same, multiple users may have the same viewport in HMDs simultaneously. {Note that the number of different viewports among users relates to the difference of users' favors.} We treat these users, who have the same viewport in HMDs, as one entity. {The number of entities is equal to the number of different viewports. To make clear the relationship between users and tiles for taking advantage of multicast, we classify the all transmitted tiles of $K$ users into $L$ viewports, and index the viewport set as ${\mathcal{L}}=\left\{{1},\cdots,{l},\cdots,{L}\right\}$}\footnote{Turning users into viewport set, we can ignore the specific unicast and multicast.}. {Each viewport has its corresponding viewport tiles and viewport users.}

In a multi-stream group $g$, we use {$\mathbf{p}_g(l,f)$ as an indicator of whether the users of viewport ${l}$ retrieve the $f$-th tile stream.} Those users with viewpoint $l$ that can retrieve tile $f$ in group $g$ are associated with
indicator $\mathbf{p}_g(l,f)${=1}; otherwise, $\mathbf{p}_g(l,f){=0}$.
The users of one viewport can only reliably receive no more than one tile simultaneously; thus, we have
\begin{align}\label{non_2}
0\leq \mathbf{p}_g(l,f_1)+\mathbf{p}_g(l,f_2)\leq 1, \quad f_1\neq f_2.
\end{align}
Note that there is no restriction among the tiles in one group according to equation \eqref{non_2}. {Selecting two tiles belonging to viewport ${l}$ as two streams into group $g$, those users with viewport $l$ can only receive one of two tiles once and retrieve the other in a new group. It causes an increase in the number of transmitted groups, which leads to the performance degradation. To solve it, each tile stream in a group belongs to a distinct viewport. To search the optimal grouping method, we extend the conception of \eqref{non_2}.
A tile $f$ in group $g$ that belongs to a viewport $l$ is associated with indicator $\mathbf{d}_g(l,f)=1$; otherwise, $\mathbf{d}_g(l,f)=0$. Thus we can write
\begin{align}\label{non}
0\leq \mathbf{d}_g(l,f_1)+\mathbf{d}_g(l,f_2)\leq 1, \quad f_1\neq f_2.
\end{align}}
Note that constraint \eqref{non_2} is a basic condition for stable transmission and constraint \eqref{non} is a necessary condition for the optimal grouping. Further, selecting more viewports into a group, which slightly increases the length of pilot sequence, can reduce the group number $G$. Empirically, the influence from the increased length of pilot sequence is relatively smaller than that from the reduced group number\footnote{The worst case is basic grouping method which has the maximum group number and the minimum length of pilot sequence.}. Also, reducing the length of pilot sequence under the minimum group number is necessary.

For a certain tile $f$, it belongs to one viewport or multiple viewports. To distinguish the two tile types, we define the former as isolated tile (IT), and the latter as coexisting tile (CT). To clarify the relationship between tile and viewport, we analyse a simple viewport set ${\mathcal{L}}=\left\{{l_1},{l_2},{l_3}\right \}$. {Denote the CTs only belonging to ${l_1}$ and ${l_2}$ as {$\xi({l_1},{l_2})$}, the CTs of ${l_3}$ as {$\xi(\sum {l_3})$}, the ITs of ${l_3}$ as {$\xi({l_3})$}, and the CTs removing ${l_2}$ from $(\sum {l_3})$ as {$\xi(\sum {l_3}-l_2)$}, respectively.} And they follow
\begin{equation}\label{belonging}
\left \{
\begin{aligned}
&\xi({l_1},{l_2},{l_3})\subseteq \xi\left(\sum {l_3}\right) \\
&\xi({l_1},{l_3})\subseteq \xi\left(\sum {l_3}-{l_2}\right) \\
&\xi({l_2},{l_3})\subseteq \xi\left(\sum {l_3}-{l_1}\right).
\end{aligned}
\right.
\end{equation}
Further, we denote the number of CTs $\xi({l_1}, {l_2})$, ITs $\xi({l_3})$, and CTs $\xi(\sum {l_3}-{l_2})$ as $\Pi({l_1}, {l_2})$, ${\mathcal{A}}({l_3})$, and $\Pi(\sum {l_3}-{l_2})$, respectively. Then, we have
\begin{align}
&\Pi({l_1},{l_3})+\Pi({l_2},{l_3})+\Pi({l_1},{l_2},{l_3})=N_p-{\mathcal{A}}({l_3})- \Pi\left(\sum {l_3}-{l_1}-{l_2}\right).
\end{align}
Based on the distribution of viewport tiles in the HMD scene, we define the viewport that has no IT as coexisting viewport (CV) and that has at least one IT as isolated viewport (IV).
\begin{figure*}[htb]
\centering
    \includegraphics[width=0.4\textwidth]{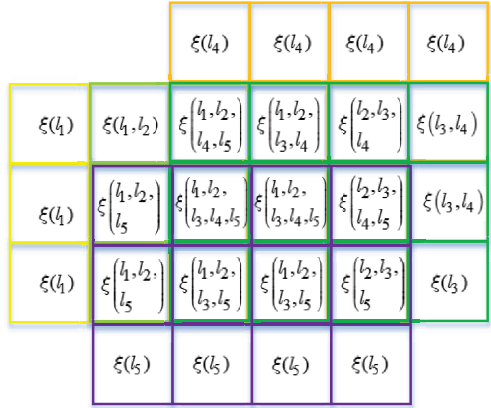}
      \caption{An example of viewports relation in multi-stream groups.}\label{relation}
\end{figure*}

For clarity, we utilize an example illustrated in Fig. \ref{relation} to elaborate the relation of multiple viewports. Each rectangular viewport has 4 tiles in the length and 3 tiles in the width, and we denote its formate as $4\times3$. We also draw the relationship among viewports in the tile region {and illustrate the relationship of equations \eqref{non} and \eqref{belonging} by the graph}. After selecting tile $\xi({l_1})$ into multi-stream group $g$, tile $\xi({l_1},{l_2})$ is unable to join in group $g$. Further, selecting tile $\xi({l_1})$ and tile $\xi(\sum{l_3}-{l_2})$ into multi-stream group $g$, each tile of viewport ${l_2}$ is unable to join in group $g$.
Normally, CVs always connect with other viewports and are unable to be selected alone, and IVs are the opposite.
Note that the goal is to select all viewport tiles into minimal groups under constraint \eqref{non}.
Hence, the focus of tile selection mainly locates on the combination between CVs.

After selecting a viewport tile into a group, the rest viewport tiles form a new graph. And the new selection in another group is processed based on the new graph. The process continues until there is no tile left.
\subsection{Optimal Multi-Stream Grouping Based on Rectangular Viewport}\label{sec:optimal_grouping}
In order to facilitate the following description, we make two definitions as follows:
\begin{definition}
Combing multiple tiles to form a multi-stream group is defined as combination, and the operation symbol is defined as $\bigcup$.
\end{definition}
\begin{definition}
The group containing all viewports is a complete group, denoted as ${\Upsilon}$.
\end{definition}

The optimal result in each selection process is picking out all viewports in the graph and form a complete group. When selecting a viewport tile into a group, whether the other viewports satisfy constraint \eqref{non} is unknown. Moreover, the combination basis for every viewport is unknown, and there is no algorithm to guarantee a complete group in each selection process. To analytically search the optimal complete groups, we firstly propose the following proposition:
\begin{proposition}\label{proposition:minimal}
Under constraint \eqref{non}, combination of rectangular viewports with identical shape $h\times v$ has the minimal complete group number $G_{\textrm{re}}=h\cdot v$, and the combination of tile $\zeta(x,y)$ satisfies
\begin{align}\label{combination}
\bigcup_{j=-\left \lfloor\frac{y}{v}\right \rfloor}^{\left \lfloor\frac{V-y}{v}\right \rfloor} \bigcup_{i=-\left \lfloor\frac{x}{h}\right \rfloor}^{\left \lfloor{\frac{H-x}{h}}\right \rfloor } \zeta(x+i\cdot h,y+j\cdot v)\overset{def}{=}{\Upsilon}
\end{align}
{where $H=\max{\{x\}}$ and $V=\max{\{y\}}$.}
\end{proposition}

\begin{IEEEproof}
See Appendix \ref{appendix}.
\end{IEEEproof}

\begin{remark}
  \emph{Proposition} \ref{proposition:minimal} firstly reveals the combination relation among all rectangular viewports in the combination problem. Secondly, \emph{Proposition} \ref{proposition:minimal} represents that rectangular viewports with identical shapes have the minimal complete group number. Applying the maximum served users in each group and non-repeated tile stream in all groups is able to make the most of multi-stream ability and minimize the length of pilot sequences in massive MIMO systems. It greatly improves the throughput and reduces delay in VR 360$^{\circ}$ video transmission. Thirdly, $G_{\textrm{re}}=h\cdot v$ well reflects the relation between transmitted group size and viewport format size.
\end{remark}
\subsection{MLMSG on Non-Rectangular Viewport}
\begin{figure*}[htb]
\centering
\subfigure[Decomposition of non-rectangular viewport ${l_1}$]
    {\includegraphics[width=0.4\textwidth]{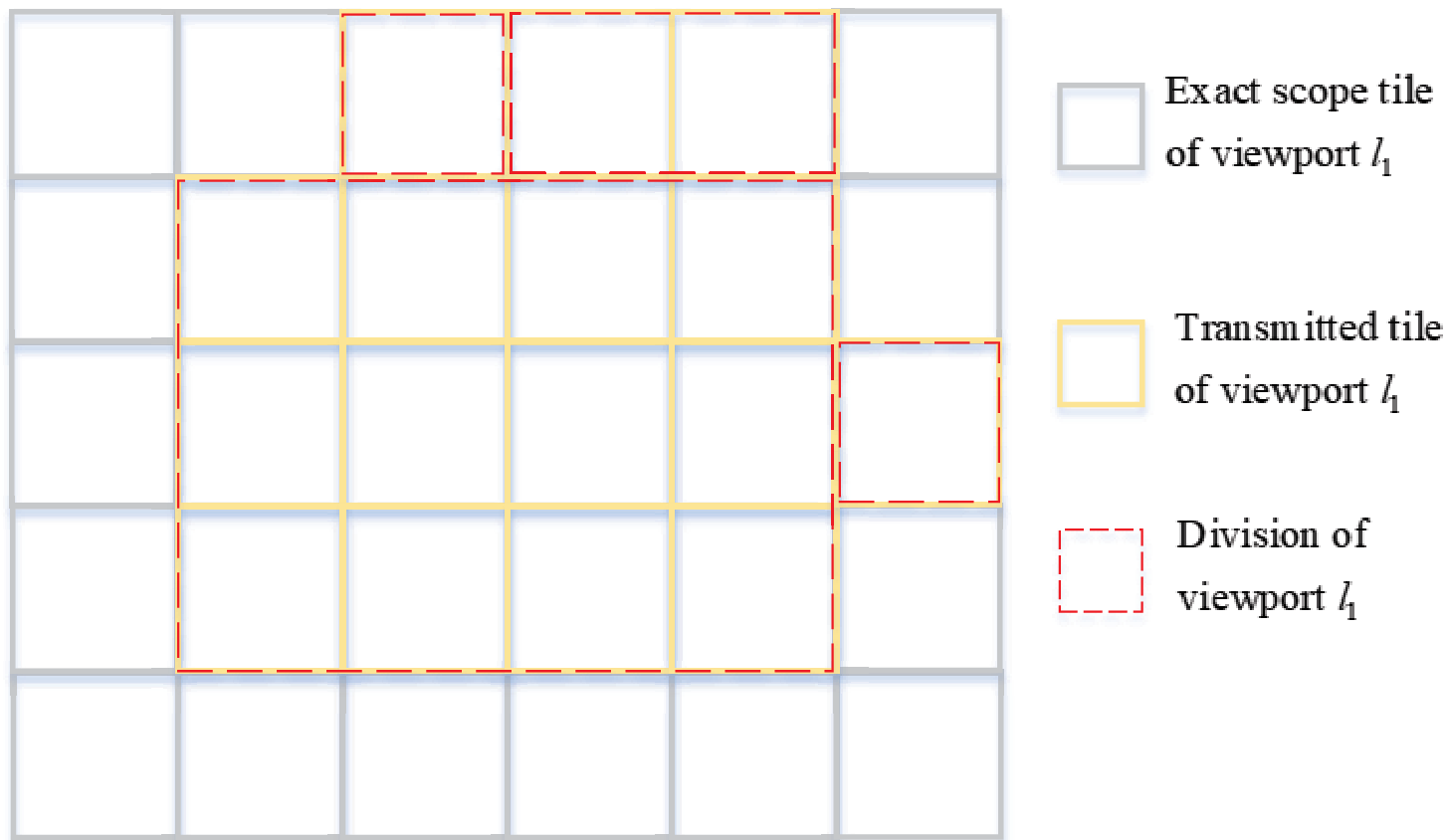}}
     \hspace{0.1in}
\subfigure[Decomposition of non-rectangular viewport ${l_2}$]
    {\includegraphics[width=0.4\textwidth]{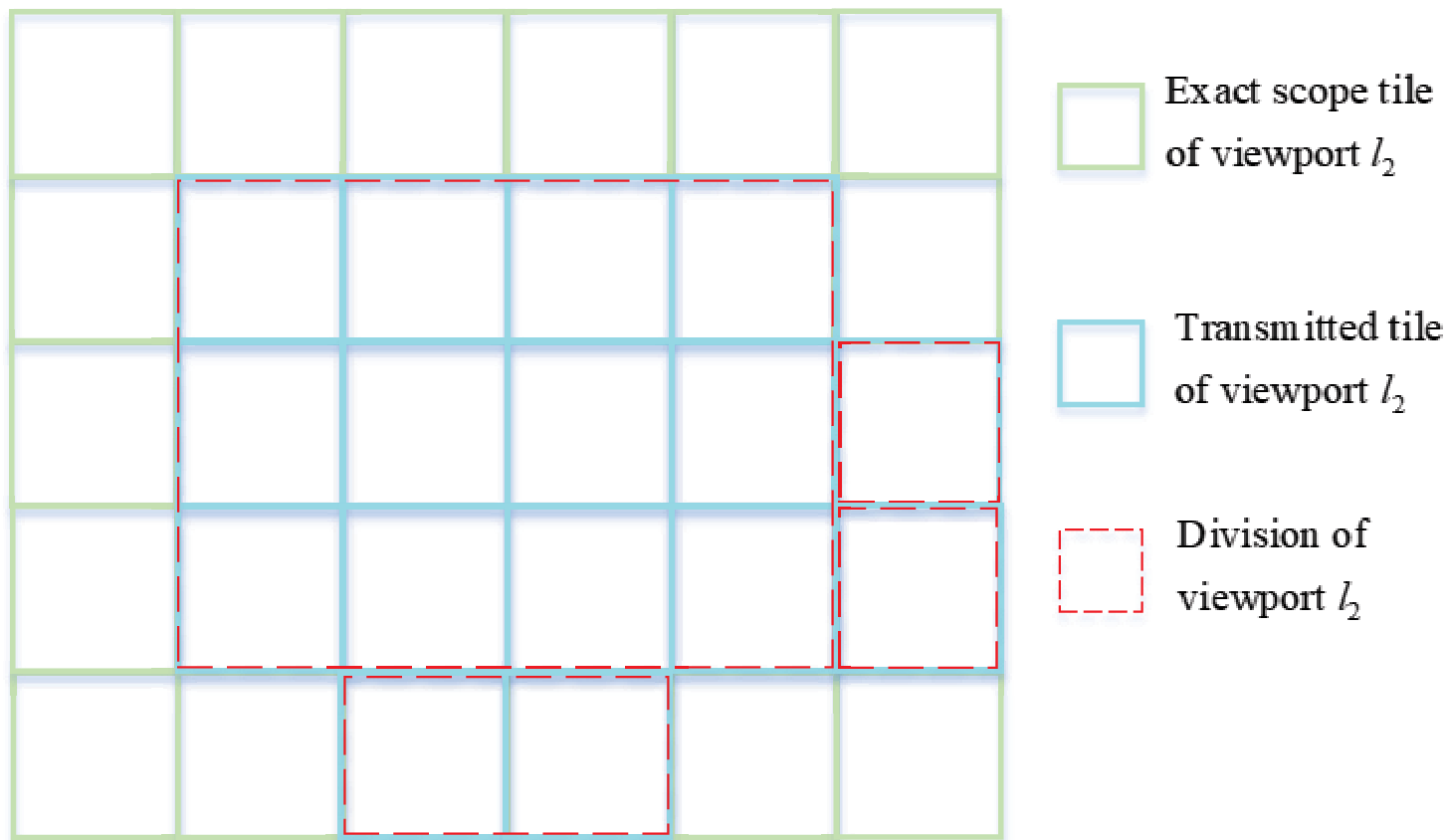}}
      \caption{Decomposition of non-rectangular viewports accordingly to their shapes.}\label{divide}
\end{figure*}

In the proposed transmission model, the transmitted viewport contains not only the predictive FoV tiles, but also other tiles in the exact scope. Thus, sometimes the shape of transmitted viewport to each user is non-rectangular. Note that the non-rectangular viewports have no minimum complete group number under constraint \eqref{non}, even though the shape of each viewport is identical.
Based on the \emph{Proposition} \ref{proposition:minimal}, we can try to structure a rectangular tile entity, i.e., decomposing the non-rectangular viewport $l$ into rectangular tile lattice $\lambda_l$ and the other rest tiles $\delta_l$\footnote{Compared with $\lambda_l$, $\delta_l$ is relative small and the combination of $\delta_l$ is only subject to the basic constraint \eqref{non_2}.}.
Then, $|\lambda_l|+|\delta_l|=N_p$. Note that the larger lattice can achieve smaller total length of pilot sequences for likely increasing the ratio of CTs to all tiles. Thus, the shape of each rectangular tile lattice is the largest shape within all non-rectangular viewports. For rectangular tile lattice $\lambda_l$, Section \ref{sec:optimal_grouping} gives the optimal multi-stream grouping. By iterative decomposition, we can decompose the rest tiles $\delta_l$ into multiple types of rectangular tile lattices according to the shape of every $\delta_l$. To meet the tile combination requirement, the quantity of each type of rectangular tile lattices for each viewport is the same. For clarity, we describe the process with an illustration given in Fig. \ref{divide}.

In Fig. \ref{divide}, the tiles in the red dashed line boxes are tile lattices. The first largest tile lattice is the central $4\times 3$ tile entity. Then, in their rest tiles, there is a lattice including two tiles with identical shapes. Last, the rest one-tile entity forms an one-tile lattice. Hence, we can decompose both non-rectangular viewports ${l_1}$ and ${l_2}$ into one central $4\times 3$ rectangular tile lattice, one $2\times 1$ rectangular tile lattice, and two $1\times1$ rectangular tile lattices, respectively.

Note that the combination in each tile lattice set is independent and the equation \eqref{combination} is still applicable for each type of tile lattice set.  Hence, the total group number $G_{\textrm{re}}$ is equal to $N_p$\footnote{When the $N_p^k$ is different for different users, the decomposition method is still applicable, and the group number $G_{\textrm{re}}$ is equal to $\max$\{$N_p^k$\}.} , and the stream number in each group is determined. The complexity of the combination of the non-rectangular viewports is linear with $N_p$, namely, $\mathcal{O}(N_p)$. Note that the MLMSG is also suitable for missing viewport tiles, and the complexity is $\mathcal{O}(N_m)$.

\section{Average QoE Maximization}\label{QoE}
The traditional retrieving method only fetches and buffers the predictive FoV tiles, which ignores the missing tiles. The missing tiles do appear in the practical scenario, which causes QoE degradation and even unacceptable VR sickness. In this section, we jointly consider the predictive transmission and the supplementary transmission for missing tiles, and explore the effect of $N_p$ on the average maximum QoE.

Without fully accurate prediction, prediction error occurs and is closely related to exceptional motion \cite{optimal_viewport}, which is caused by the multi-stimuli in HMDs. Researchers in \cite{VRSA} leverage realism loss and reconstruction loss to predict the intensity of exception motion of fragment frames, which can provide experimental evidence on the exact scope. Hence, the size of exact scope tiles is a function of the stimuli, which can be obtained based on the previous prediction model. Generally, we assume that the request probability of each tile in the exact scope is equal. Thus, the expected number of missing tiles $N_m$ is
\begin{align}\label{J}
N_m=\left\lceil M\cdot \frac{S-N_p}{S} \right\rceil
\end{align}
where $M$ and $S$ are the tiles number of FoV format and exact scope format, respectively.

According to the expression of achievable SE and multi-stream grouping,
\begin{equation}\label{GJ}
\left \{
\begin{aligned}
&G=N_p \\
&J=N_m.
\end{aligned}
\right.
\end{equation}
Hence, $N_p$ is also an optimization variable that determines $G$ and $J$. Thus, we reformulate problem (\textbf{P0}) into problem (\textbf{P1})

\begin{align}
\left(\textbf{P1}\right) \quad \max_{\eta_p,\eta_m, N_p}\quad & \frac{1}{K}\sum_{k=1}^{K}\alpha_k \eta_p-\beta_k N_m(\eta_p-\eta_m) \nonumber \\
\textrm{s.t.}\quad & \eqref{pre_com}, \eqref{RD}, \eqref{J}, \eqref{GJ} \nonumber\\
& \sum_{g=1}^{G}\frac{T_{1}\eta_p}{v_g}+\sum_{j=1}^{J} \frac{(T_{1}-T_{2})\eta_m}{v_j}\leq T_{1} \label{time_all_2} \\
& \sum_{j=1}^{J} \frac{T_{2}\eta_m}{v_j}\leq T_y \label{time_sick_2}
\end{align}
where $v_g$ and $v_j$ are the transmission rates of predictive group $g$ and missing group $j$, respectively. Note that $v_g=W\cdot{{{\Gamma}}}_{g,\epsilon}$ is achieved in the predictive transmission but $v_j$ is unavailable without knowning specific tile and user.

To solve the unknown $v_j$ and meet the tolerant latency constraint, $v_j$ is suggested to be a valid value, which is little bit smaller than the actual value. For simplicity, the missing groups reuse the definition of predictive groups,
{i.e., $\mathbb{J}_j$, $\mathbb{J}_j(f,b)$, $\mathbb{j}_{j,f}$, $\boldsymbol{\Psi}_j(f,b)$, $\mathbf{U}_j(f,b)$, and $\mathbf{A}_j(f,b)$ have similar meanings to $\mathbb{J}_g$, $\mathbb{J}_g(f,b)$, $\mathbb{j}_{g,f}$, $\boldsymbol{\Psi}_g(f,b)$, $\mathbf{U}_g(f,b)$, and $\mathbf{A}_g(f,b)$, respectively. The differences are that group indices $g$ turn to be $j$ for missing group $j$.}

In terms of transmission rate in $j$-th missing group by MRT precoding, {the SINR of each stream is the same} and we formulate it as
\begin{align}\label{MRT}
{{\Omega}}^{\textrm{MRT}}_{j,\epsilon}=\frac{P}{\sum_{i=1}^{F}\frac{1}{\min_{b \in {\mathbb{j}_{j,i}}}{(\mathbf{A}^{\textrm{MRT}}_j(i,b))}}} &=\frac{P}{\sum_{i=1}^{F}\frac{1}{\min_{b\in {\mathbb{j}_{j,i}}}\left[\frac{N\sigma_jq^u({\boldsymbol{{\Psi}}}_j(i,b))^2}{\left(1+\sum_{t=1}^{B} \sigma_jq^u{\boldsymbol{{\Psi}}}_j(i,t)\right)\left(1+{\boldsymbol{{\Psi}}}_j(i,b)P\right)}\right]}} \nonumber \\
&=\frac{NP}{\sum_{i=1}^{F}\max_{b\in  {\mathbb{j}_{j,i}}}\left [\frac{\left(1+\sum_{t=1}^{B}\sigma_jq^u{\boldsymbol{{\Psi}}}_j(i,t)\right)\left(1+{\boldsymbol{{\Psi}}}_j(i,b) P\right)}{\sigma_jq^u({\boldsymbol{{\Psi}}}_j(i,b))^2}\right]}
\end{align}
where $\frac{1+{\boldsymbol{{\Psi}}}_j(i,b)P}{\sigma_gq^u({\boldsymbol{{\Psi}}}_j(i,b))^2}$ decreases monotonously as ${\boldsymbol{{\Psi}}}_j(i,b)$ increases. In the worst case, the minimal large-scale fading coefficient in each stream is the same as the minimal one of the total users. Hence,
\begin{align} \label{SINR_de}
{{\Omega}}^{\textrm{MRT}}_{j,\epsilon} \geq \frac{NP\cdot \frac{\sigma_jq^u {{{\Psi}}}^2_{\textrm{min}}}{1+{{{\Psi}}}_{\textrm{min}}P}}{\sum_{i=1}^{F} 1+\sum_{t=1}^{B}\sigma_jq^u
{{{\Psi}}}_j(i,t)}
= NP \cdot \frac{\frac{\sigma_jq^u{{{\Psi}}}^2_{\textrm{min}}}{1+{{{\Psi}}}_{\textrm{min}}P}}
{\sigma_j+\sigma_jq^u{{{\Psi}}}_{\textrm{all}}} =NP\cdot\frac{\frac{q^u{{{\Psi}}}^2_{\textrm{min}}}{1+{{{\Psi}}}_{\textrm{min}}P}}
{1+q^u{{{\Psi}}}_{\textrm{all}}}
\end{align}
where ${{{\Psi}}}_{\textrm{all}}=\sum_{i=1}^{F}\sum_{t=1}^{B}{\boldsymbol{{\Psi}}}_j(i,t)$, and ${{{\Psi}}}_{\textrm{min}}$ is the minimal value of the user large-scale fading coefficients. Denote the right term in \eqref{SINR_de} as ${{\Omega}}^{\textrm{MRT}}_{j,\textrm{min}}, j\in \mathcal{J}$, which has no association with the length of pilot sequence and group tile index.

Likewise, we formulate the SINR of $j$-th group by ZF precoding as
\begin{align}\label{ZF}
{\Omega}^{\textrm{ZF}}_{j,\epsilon}=\frac{P}{\sum_{i=1}^{F}\frac{1}{\min_{b \in  {\mathbb{j}_{j,i}}}({\mathbf{A}}^{\textrm{ZF}}_j(i,b))}}
&=\frac{P}{\sum_{i=1}^{F}\frac{1}{\min_{b\in  {\mathbb{j}_{j,i}}}\left[\frac{(N-\sigma_j) \cdot {\mathbf{U}}_j(i,b)}{1+\left({\boldsymbol{{\Psi}}}_j(i,b)-{\mathbf{U}}_j(i,b)\right)P}\right]}} \nonumber \\
&=\frac{P}{\sum_{i=1}^{F}\max_{b\in {\mathbb{j}_{j,i}}}\left [\frac{1+\left({\boldsymbol{{\Psi}}}_j(i,b)-{\mathbf{U}}_j(i,b)\right)P}{(N-\sigma_j)\cdot {\mathbf{U}}_j(i,b)}\right]}
\end{align}
where $\frac{1+{\boldsymbol{{\Psi}}}_j(i,b)P}{{\mathbf{U}}_j(i,b)}$ decreases monotonously as ${\boldsymbol{{\Psi}}}_j(i,b)$ increases. The worst case is that the minimal large-scale fading coefficient in each stream is ${{{\Psi}}}_{\textrm{min}}$. Similarly,
\begin{align}\label{SINR_ZF}
{{\Omega}}_{j,\epsilon}^{\textrm{ZF}}\geq\frac{\left(N-\sigma_j\right)q^u{{{\Psi}}}^2_{\textrm{min}}P}{1+q^u{{{\Psi}}}_{\textrm{all}}+ \left(1+q^u{{{\Psi}}}_{\textrm{all}}\right){{{\Psi}}}_{\textrm{min}}P-\sigma_jq^u{{{\Psi}}}^2_{\textrm{min}}P}
\end{align}
where the right term is denoted as ${{\Omega}}_{j,\textrm{min}}^{\textrm{ZF}}$.
The approximate errors in MRT precoding and ZF precoding are extremely small, which are revealed in Section \ref{simulation}.

{We denote the minimum SEs of group $j$  in MRT precoding and ZF precoding by ${\Gamma}_{j,\textrm{min}}^{\textrm{MRT}}=(1-\sigma_j/T)\textrm{log}_2\left(1+{{\Omega}}_{j,\textrm{min}}^{\textrm{MRT}}\right)$ and ${\Gamma}_{j,\textrm{min}}^{\textrm{ZF}}=(1-\sigma_j/T)\textrm{log}_2\left(1+{{\Omega}}_{j,\textrm{min}}^{\textrm{ZF}}\right)$, respectively.
Further, $\sigma_j, \forall j \in \mathcal{J}$, is unknown. In the worst case, the tile lattices in the multi-stream grouping are all $1\times1$ tile lattices; thus $\sigma_j$ is equal to $L$ in ${\Gamma}_{j,\textrm{min}}^{\textrm{MRT}}$ and ${\Gamma}_{j,\textrm{min}}^{\textrm{ZF}}$.
Hence, the results $v_j$ of problem (\textbf{P1}) in MRT precoding and ZF precoding approximate to $v_{j}^{\textrm{MRT}}=W\cdot{{{\Gamma}}}_{j,\textrm{min}}^{\textrm{MRT}}$ and $v_{j}^{\textrm{ZF}}=W\cdot{{{\Gamma}}}_{j,\textrm{min}}^{\textrm{ZF}}$, respectively. Note that the approximation is also able to be applied to predictive tile groups.}

By analysing, $\eta_p$ and $\eta_m$ in constraints \eqref{time_all_2} have linear relationship and are also linear with $N_p$, and $\eta_m$ in \eqref{time_sick_2} is linear with $N_m$, whereas the product of $N_m$ and $(\eta_p-\eta_m)$ makes the optimization objective nonlinear. To make it linear, the general method is to set $N_m$ or $(\eta_p-\eta_m)$ as a constant. $N_p$, the set of which is $\left\{M,\cdots,S\right\}$, decides $N_m$, the cardinality of which is much smaller than that of $(\eta_p-\eta_m)$. Thus, we set $N_p$ as a constant from $M$ to $S$, {and respectively denote $\eta_p(N_p)$ and $\eta_m(N_p)$ as the encoding rates.} The final QoE score of the optimization objective is the maximal value among the calculated results.

\begin{figure*}[htb]
\centering
    \includegraphics[width=0.4\textwidth]{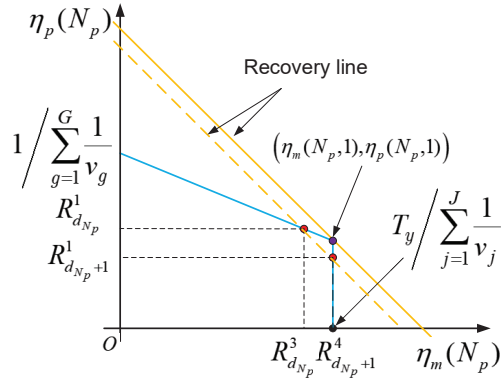}
      \caption{The recovery policy of two-dimensional integer point.}\label{point}
\end{figure*}

Note that the two discrete variables $\eta_p$ and $\eta_m$ make each calculation non-convex. Based on the general linear programming method, we relax discrete variables into continuous variables and recover them from the optimization solution. Hence, with fixed $N_p$, {we relax $\eta_p(N_p), \eta_m(N_p) \in \left[R_1,R_D\right]$}, and problem (\textbf{P1}) in the calculation turns to be an ILP problem (\textbf{P2}):
\begin{align}
\left(\textbf{P2}\right) \quad \max_{\eta_p{(N_p)},\eta_m{(N_p)}}\quad & \frac{1}{K}\sum_{k=1}^{K}\alpha_k \eta_p{(N_p)}-\beta_k N_m\left[\eta_p{(N_p)}-\eta_m{(N_p)}\right] \nonumber \\
\textrm{s.t.}\quad &\eqref{pre_com},\eqref{J}, \eqref{GJ}, \eqref{time_all_2}, \eqref{time_sick_2} \nonumber\\
& \eta_p{(N_p)}, \eta_m{(N_p)} \in \left[R_1,\cdots,R_d,\cdots,R_D\right]\label{RD_2}
\end{align}
which can be efficiently solved through the convex optimization toolbox \cite{CVX}. In the fixed $N_p$ calculation, we obtain {the optimization solutions $\eta_p(N_p,1)$ and $\eta_m(N_p,1)$}, which locate in intervals $\left[R_{d_{N_p}}^1,R_{d_{N_p}+1}^1\right]$ and $\left[R_{d_{N_p}}^2,R_{d_{N_p}+1}^2\right]$, respectively. According to the two-dimensional linear programming, the closest two-dimensional integer point is only relevant to {$\eta_p(N_p,1)$}, which is illustrated in Fig. \ref{point} for clarity.
Thus, the two variables recovery starts from variable {$\eta_p(N_p,1)$}, which is either $R_{d_{N_p}}^1$ or $R_{d_{N_p}+1}^1$. For fixed $R_{d_{N_p}}^1$ and $R_{d_{N_p}+1}^1$, {we respectively obtain the optimal encoding rates of missing tiles and indicate them by $R_{d_{N_p}}^3$ and $R_{d_{N_p}+1}^4$}, {and we use $\mathcal{Q}(N_p,R_{d_{N_p}}^1,R_{d_{N_p}}^3)$ and $\mathcal{Q}(N_p,R_{d_{N_p}+1}^1,R_{d_{N_p}+1}^4)$ to indicate their QoE scores, respectively}. Hence, we easily obtain the recovery policy and QoE score in the fixed $N_p$ calculation as follows
{\begin{align}
{\mathcal{Q}}_{N_p}=\max \left \{ {\mathcal{Q}}(N_p,R_{d_{N_p}}^1,R_{d_{N_p}}^3) , {\mathcal{Q}}(N_p,R_{d_{N_p}+1}^1,R_{d_{N_p}+1}^4)\right \}.
\end{align}}
Therefore, the optimal QoE is
\begin{align}
{\mathcal{Q}}_{\textrm{op}}=\max\{{\mathcal{Q}}_S,\cdots,{\mathcal{Q}}_M\}.
\end{align}
Therefore, we accordingly obtain the optimal $N_p$, $\eta_p$, and $\eta_m$. In addition, the linear programming makes the complexity as $(S-M)\mathcal{O}(K)$, which is a low value.
\section{Simulation Results}\label{simulation}
In the section, we {run simulation in Matlab} to verify the performance of MLMSG and the joint consideration between predictive tiles and missing tiles in QoE driven VR $360^{\circ}$ video massive MIMO transmission.
\subsection{Simulation Setup}
We consider VR transmission in a single cell where the radii $r1$ and $r2$ are 45 {meters} and 40 {meters}, respectively. The number of users $K$ {and} antennas $N$ are 100 and 128, respectively.
We model the large-scale fading coefficient for user $k$ as ${\psi_k}={c}/\tau^{\kappa}_{k}$, where $\tau_k$ is the distance between user $k$ and the BS, $\kappa=3.76$ is the pass-loss exponent, and ${c}=10^{-3.5}$ is a constant \cite{UM}. We set the transmission bandwidth $W$ as {100} MHz at a carrier frequency of 2 GHz. The coherence bandwidth and coherence time are 200 kHz and 1 ms, respectively, which contribute to the coherence interval of 200 symbols.
We set the noise power spectral density, HMD power $P_{{u}}$, and total downlink power $P_{d}$ as $\sigma^2=-174$ dBm/Hz, $0.1$ Watts, and 10 Watts, respectively. Thus, the normalized uplink power and total downlink power are $q^{{u}}=P_{{u}}/(W\cdot \sigma^2)$ and $P=P_{d}/(W\cdot \sigma^2)$, respectively.

{Given that the prediction accuracy decreases sharply as the interval time $T_1$ increases, $T_1$ is usually set at 200 \textrm{ms} for high prediction accuracy \cite{Flare}, \cite{predictive_360}. Further, we consider a moderate $T_2$ to balance transmission time with computing time and rendering time. Thus, we set the time interval of the predictive tiles and the missing tiles to $T_1=200$ $\textrm{ms}$ and $T_2=90$ $\textrm{ms}$, respectively.} The authors \cite{mag_VR} recommend the tolerant latency of VR sickness to be $T_y=10$ $\textrm{ms}$. In the simulation, the equirectangular format of VR 360$^{\circ}$ video by tilling projection is $12 \times 12$, and FoV format is $5 \times 4$. {The prediction accuracy is about 90$\%$ for $T_1=200$ \textrm{ms} \cite{Flare}, and the exact scope is a litter larger than the FoV. Thus, we consider three reasonable exact scope formats: $6\times4$, $5\times5$, and $6\times5$.} Further, we define the interval of encoding rate of each tile as $10^5$ bps.
{As for the weights of the major tiles quality and the perceptual difference, we adopt the concept in \cite{weight_2} and consider a moderate variation. Thus, set $\alpha_k=[1.9,2.1], \beta_k=[\bar{\beta}-0.02,\bar{\beta}+0.02]$, where $\bar{\beta}=\{0.1,0.2,\cdots,1\}$.} For clarity, we summarize our simulation parameters in Table \ref{table:parameter}.
\begin{table}[htb]
\caption{Simulation Parameters}\label{table:parameter} \centering{
\scalebox{0.8}{\begin{tabular}{|l|l||l|l|}\hline \textbf{Parameter}
& \textbf{Value} & \textbf{Parameter} & \textbf{Value}\\ \hline
\rule{0pt}{12pt}$N$ &128 & $K$ & 100 \\ \hline
\rule{0pt}{12pt}$T$ &200 & $T_1$ & 200 ms \\ \hline
\rule{0pt}{12pt}$T_2$ &90 ms & $T_y$ & 10 ms \\ \hline
\rule{0pt}{12pt}$\alpha_k$& [1.9,2.1]& $\bar{\beta}$ &$\{0.1,\cdots,1\}$ \\ \hline
\rule{0pt}{12pt}$r_1$ &40 & $r_2$ & 45 \\ \hline
\rule{0pt}{12pt}$\sigma^2$ & -174 dBm/Hz  & ${c}$ & $10^{-3.5}$ \\ \hline
\rule{0pt}{12pt}$\kappa$ & 3.76 &$\tau_k$ & $[r_1,r_2]$  \\ \hline
\rule{0pt}{12pt}$P_{{u}}$ & 0.1 W& $P_{{d}}$ & 10 W \\ \hline
\rule{0pt}{12pt}$W$ & 100 MHz & Equirectangular format & $12\times 12$ \\ \hline
\rule{0pt}{12pt}FoV format & $5\times 4$ & Encoding rate interval & $10^5$ bps\\ \hline
\end{tabular}}}
\end{table}
\subsection{Performance Evaluations and Comparisons}
In this subsection, we show the performance of the proposed MLMSG and adjustment of value $N_p$ in QoE driven VR $360^{\circ}$ video massive MIMO transmission. To the best of our knowledge, there has been no previous work proposing a complete system for VR $360^{\circ}$ video massive MIMO transmission. The basic grouping (BG) mode {derived from uni-stream multicast in \cite{multicast_l}, is to count} the indices of all users' tiles and transmitting each tile by uni-stream multicast. And the previous works on VR $360^{\circ}$ video transmission fix value $N_p$ as the FoV size $M$. Hence, we evaluate and compare the proposed methods in two parts: MLMSG and BG, with variable $N_p$ (VN) and fixed $N_p=M$ (FN), respectively.

\begin{figure*}[htb]
\centering
\subfigure [Average SINR versus $N_m$]
    {\includegraphics[width=0.4\textwidth]{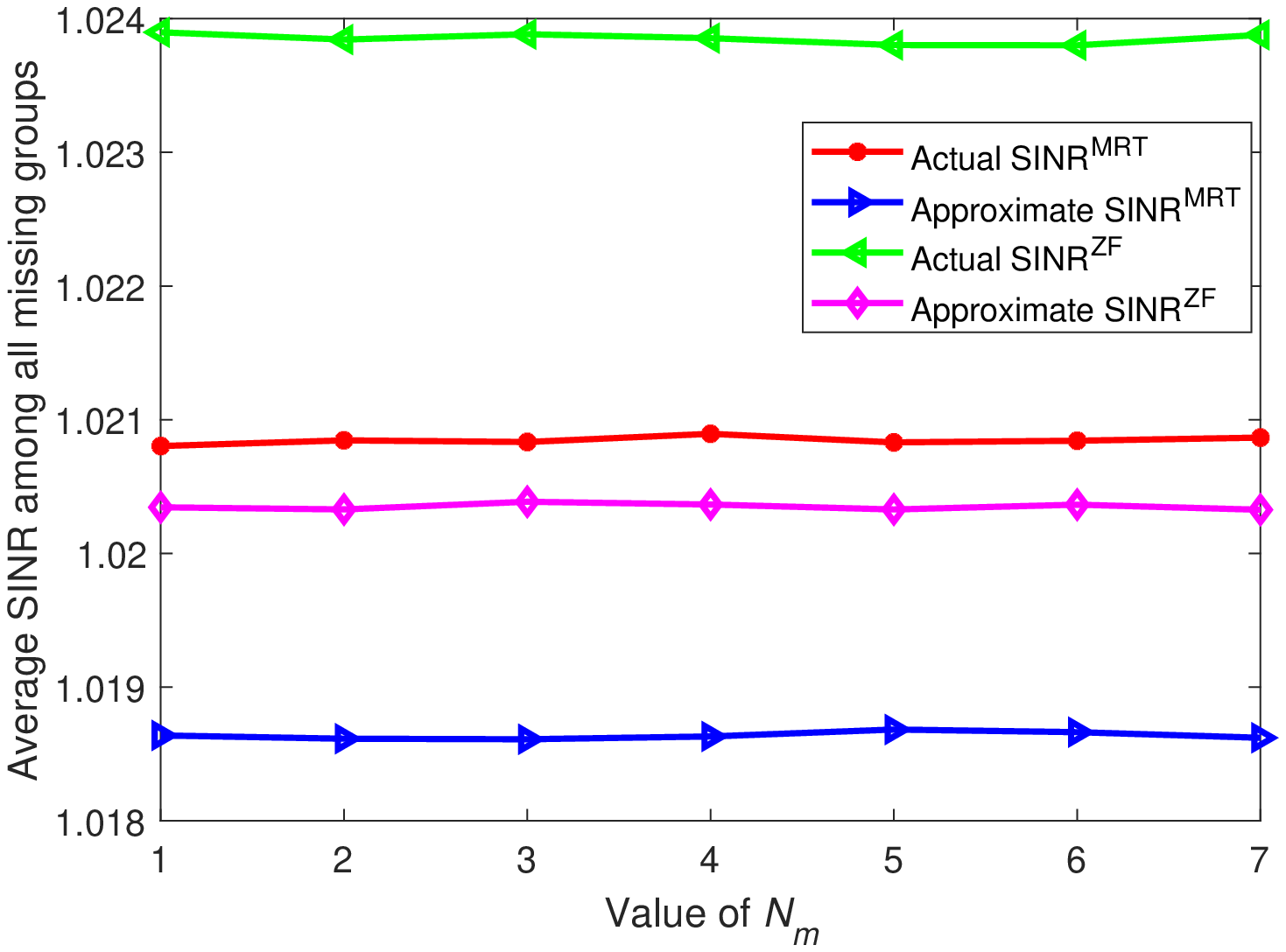}\label{fig:SINR}}
\hspace{0.1in}
\subfigure [Average SE versus $N_m$]
    {\includegraphics[width=0.4\textwidth]{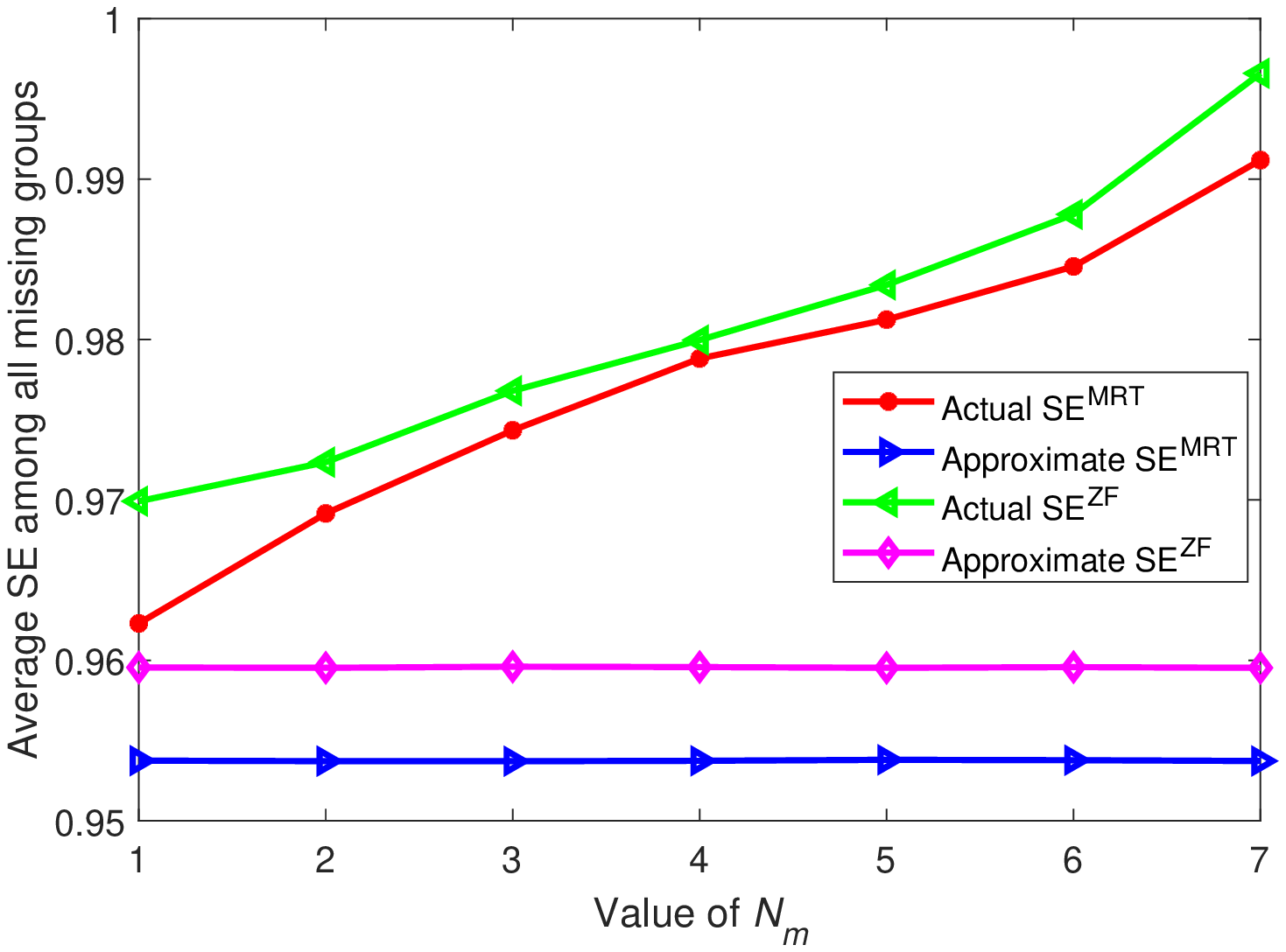}\label{fig:SE}}
      \caption{The comparison between approximate value and actual value.}\label{fig:approximate}
\end{figure*}

To evaluate the approximate processing of $v_j$ described in Section \ref{QoE}, we illustrate the actual value and approximate value versus $N_m$ in Fig. \ref{fig:approximate}. We select the exact scope format to be $6\times 5$; thus the expected number of missing tile groups $J=N_m$ can change from 1 to 7. In Fig. \ref{fig:SINR}, the approximate SINR is extremely close to the actual SINR, and the approximate error is about 0.2$\%$. {The approximate SINR and actual SINR by ZF precoding are little larger than those by MRT precoding, respectively.} In Fig. \ref{fig:SE}, the average actual SE increases with $N_p$. The reason is that some tile lattices contain more CTs such that the total pilot sequences turn smaller. The approximation error between the worst approximate SE and the actual SE in missing groups is small and the largest approximation error is 3.5$\%$, which is feasible for performance guarantee.

To evaluate and compare MLMSG and BG, we leverage the transmission delay caused by transmitting one bit of each tile on unit bandwidth, and denote it as $\rho$, which is unrelated to the encoding rate. By MLMSG method, we obtain the approximate value of $v_j$ in advance.
But the unknown value $\sum_{t=1}^{B}\sigma_jq^u{\boldsymbol{{\Psi}}}_j(f,t)$ and the variable value $J$ make value $v_j$ in BG method fluctuates over a wide range, which makes it hard to estimate the approximate value of $v_j$. Thus, we mainly calculate the value of predictive tiles, $\rho=\sum_{g=1}^{G}1/{{{\Gamma}}}_{g,\epsilon}$, which well reflects the performances of MLMSG and BG. According to the FoV format $5\times 4$ and  three considered exact scope formats, we present the results of $\rho$ with predictive tile formats of $5\times4$, $6\times4$,  $5\times5$, and $6\times5$ in Table \ref{table:delay_per}. For this scenario, MRT and ZF have similar behaviors. The value $\rho$ of MLMSG is much smaller than that of BG in all four tile formats, and the reduction is about 23 percent. Compared with BG, the difference is that MLMSG makes the most of the multi-stream ability in MIMO systems and the tile-grouping to reduce group number and length of pilot sequences. It validates that utilizing the multi-stream ability of massive MIMO systems and the optimal multi-stream grouping can greatly improve the throughput and reduce delay in VR 360$^{\circ}$ video transmission.
\begin{table}[htb]
\caption{Experimental Results ($\rho$)}\label{table:delay_per} \centering{
\scalebox{0.8}{\begin{tabular}{|l|l|l|l|l|}\hline
\rule{0pt}{15pt}Transmission mode& $5\times 4$ & $6\times 4$ & $5\times 5$ & $6\times 5$\\ \hline
\rule{0pt}{15pt}$\textrm{MLMSG}^{\textrm{MRT}}$ &  19.7507 & 23.8946& 24.5218&29.6776 \\ \hline
\rule{0pt}{15pt}$\textrm{BG}^{\textrm{MRT}}$ & 26.2308 & 31.0087& 31.7641 &37.4574\\ \hline
\rule{0pt}{15pt}$\textrm{MLMSG}^{\textrm{ZF}}$ &19.7411  & 23.8876 & 24.5080 &29.6666\\ \hline
\rule{0pt}{15pt}$\textrm{BG}^{\textrm{ZF}}$ &26.1884  & 30.985 & 31.7241 &37.4193\\ \hline
\end{tabular}}}
\end{table}

\begin{figure*}[htb]
\centering
\subfigure [Exact scope format 6$\times$4]
    {\includegraphics[width=0.32\textwidth]{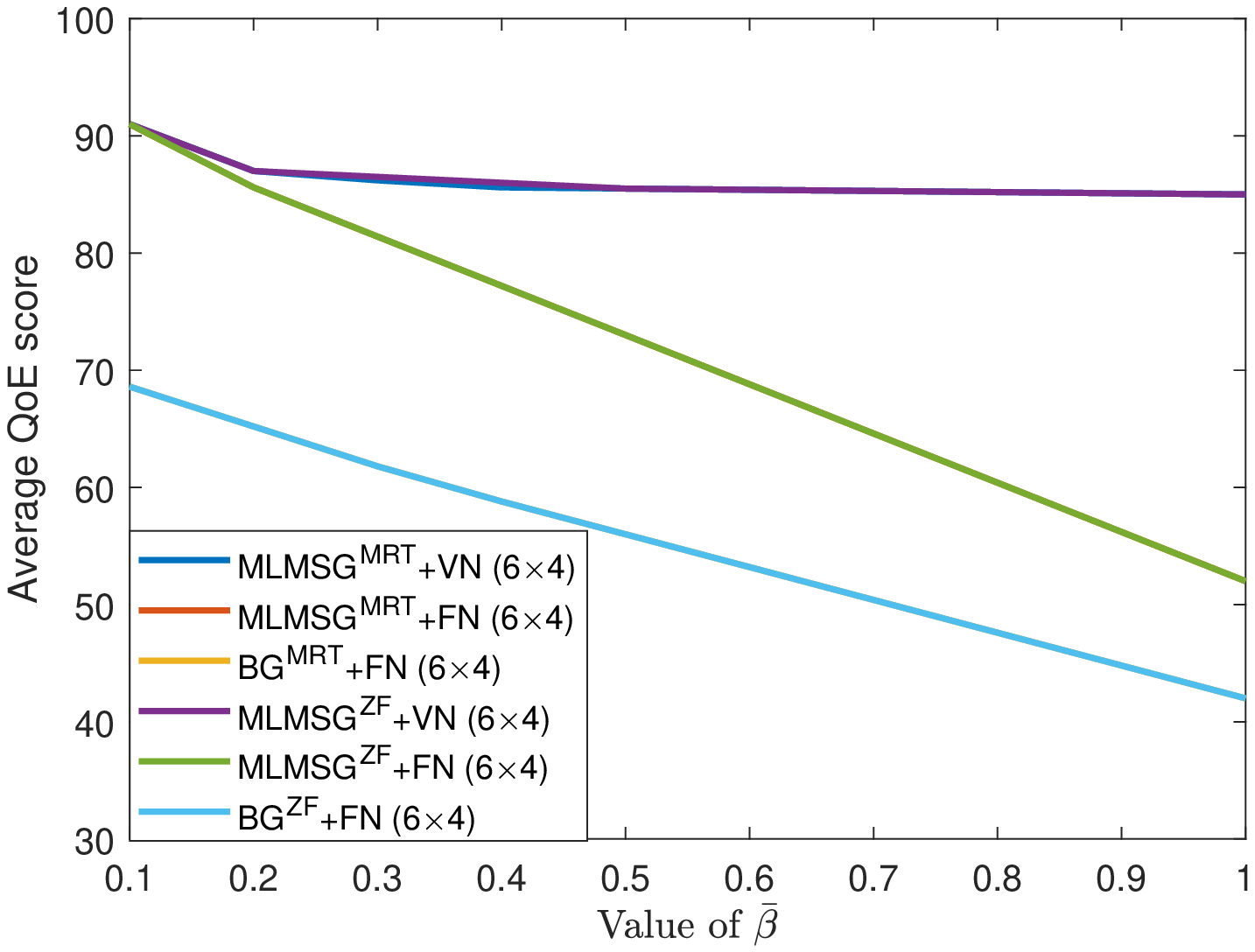}\label{fig:64}}
\subfigure [Exact scope format 5$\times$5]
    {\includegraphics[width=0.32\textwidth]{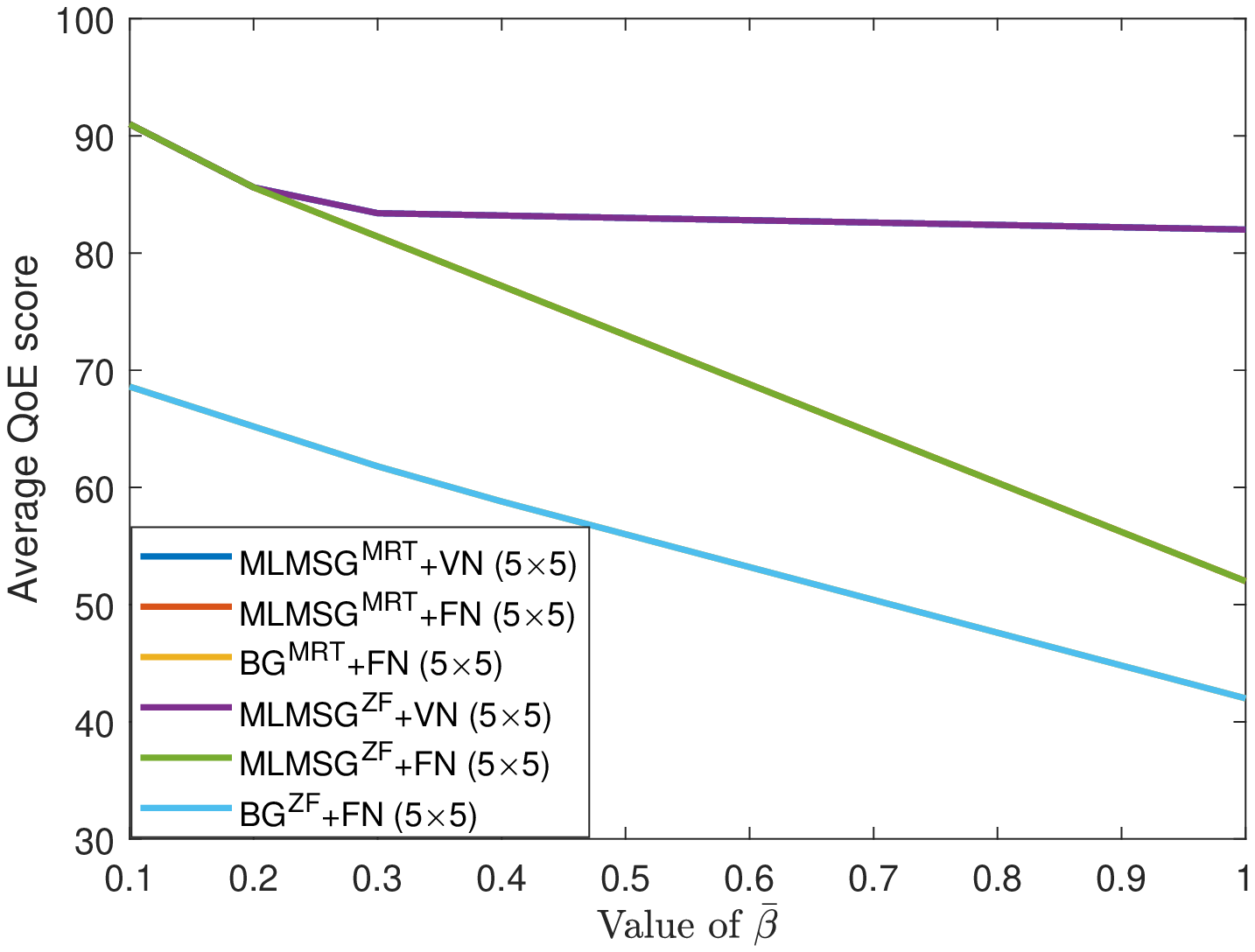}\label{fig:55}}
\subfigure [Exact scope format 6$\times$5]
    {\includegraphics[width=0.32\textwidth]{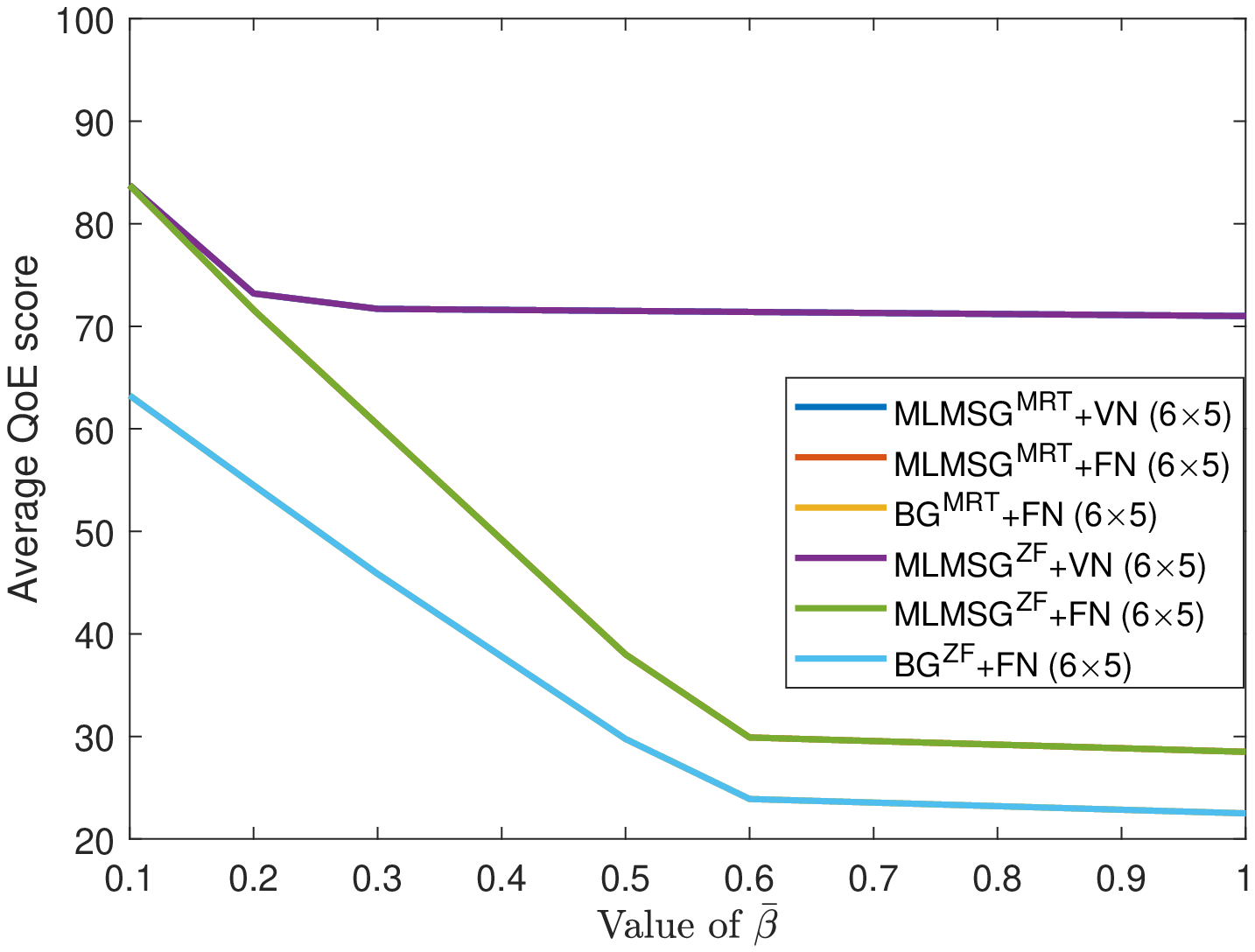}}
      \caption{Average QoE score versus value of $\bar{\beta}$ for different algorithms in three different exact scope formats.}\label{fig:QoE}
\end{figure*}

To analyze the performance of average QoE score, we mainly evaluate and compare the combination of MLMSG and VN (MLMSG$+$VN), the combination of MLMSG and FN (MLMSG$+$FN), and the combination of BG and FN (BG$+$FN). Fig. \ref{fig:QoE} presents the average QoE scores of MLMSG$+$VN, MLMSG$+$FN, and BG$+$FN.
The horizontal axis is the value of $\bar{\beta}$ and the vertical axis is the average QoE score. We restate that higher saliency of multiple stimuli leads to higher $\bar{\beta}$ for user experience.
Normally, the average QoE scores of all three methods descend as $\bar{\beta}$ increases. Benefiting from the high throughput, MLMSG can significantly improve the QoE score compared with BG. And MLMSG$+$VN has a better performance than MLMSG$+$FN when $\bar{\beta}$ turns larger. For VN, BS adjusts the encoding rate of predictive tiles to reduce the value of penalty factor while the major tiles quality descends. For FN, the penalty factor increases rapidly without adjustment, which leads to the rapid descent of the final QoE score. When the exact scope turns larger, the QoE scores of three algorithms turn smaller for more transmitted tiles and lower encoding rate. {Further, the results shows that the difference between MLMSG+VN and MLMSG+FN in Fig. \ref{fig:64} and Fig. \ref{fig:55} is detectable, though difference between 24 and 25 is little small. It demonstrates that width and height have a little influence on QoE performance.} Also, the value of $\bar{\beta}$, which distinguishes MLMSG$+$VN and MLMSG$+$FN, turns smaller for the larger exact scope format. For the exact scope with format 6$\times$5 and moderate $\bar{\beta}$, which are closer to the real multi-stimuli VR video, the average QoE score of MLMSG$+$VN maintains an acceptable level but that of MLMSG$+$FN decreases sharply as $\bar{\beta}$ increases. It validates that VN aimed at real supplementary transmission of missing tiles can improve and guarantee the QoE score.

\section{Conclusions and Future Work}\label{conclusion}
In this paper, we have investigated the problem of maximizing the average QoE in VR 360$^\circ$ video massive MIMO transmission. Based on the investigations of the previous works, we considered a practical scenario, and proposed a stable transmission scheme according to the supplementary transmission for missing tiles and unacceptable VR sickness. The integer variables and their relation make the average QoE objective be formulated as an INLP probelm. Leveraging the derived expression of the achievable SE of each tile group, we proposed the MLMSG$+$VN algorithm, and turned the INLP problem into an ILP problem by fixing the quantity of predictive tiles. With variables relaxation and recovery, we finally achieve the optimal average QoE. Simulation results suggest that our proposed MLMSG$+$VN algorithm, with pretty low complexity, improves and guarantees VR 360$^{\circ}$ video QoE. Further, the large improvement validates that the massive MIMO systems with the characteristics of the high overall throughput and the multi-stream ability are very suitable for VR 360$^{\circ}$ video transmission.

{In addition, 360$^{\circ}$ VR motion sickness, a sensory mismatch between the vestibular system and the visual system, is another challenging issue. The desired scene in the FoV and interactive
virtual world should be presented immediately and satisfactorily during user motion, which demands high-rate network links. Given the low latency and high transmission rate of massive MIMO, adapting our method to head motion to improve user experience is an important future direction.}
\appendices

\section{Proof of Proposition \ref{proposition:minimal}}\label{appendix}
\begin{figure*}[htb]
\centering
\subfigure[Horizontal MVC case]
    {\includegraphics[width=0.35\textwidth]{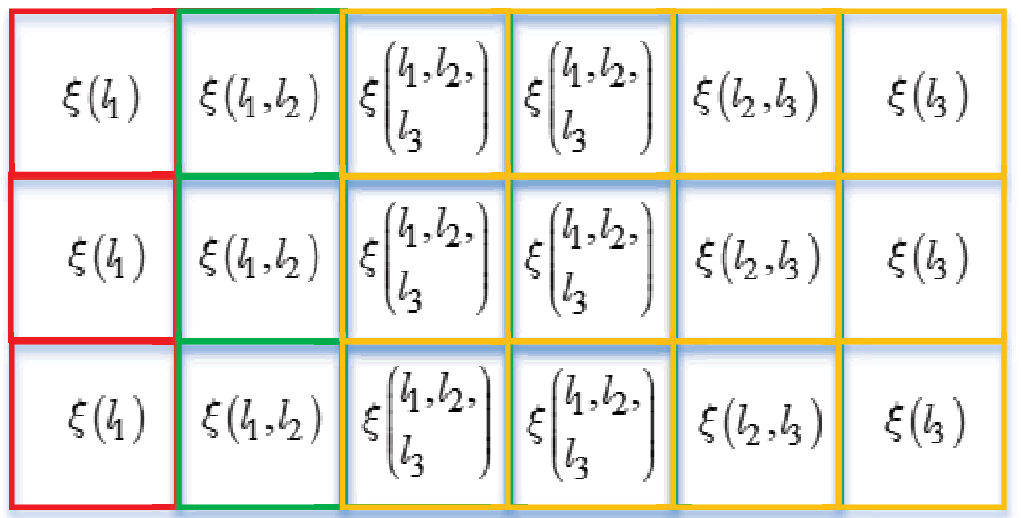}\label{three_1}}
 \hspace{0.1in}
\subfigure[Vertical MVC case]
    {\includegraphics[width=0.25\textwidth]{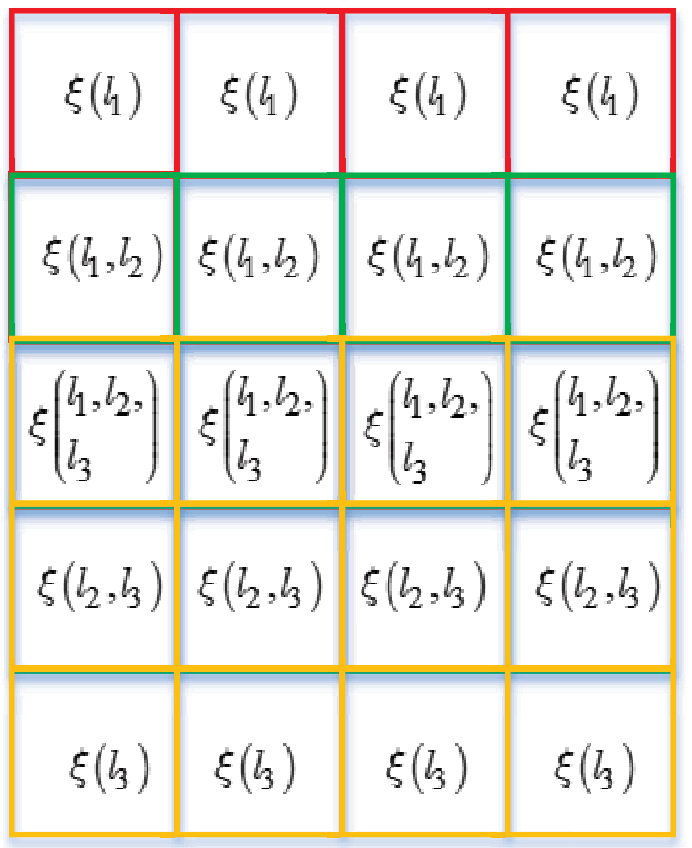}\label{three_2}}
    \caption{Two basic cases of three $4\times 3$ viewports.}\label{three}
\end{figure*}

To prove \emph{Proposition} \ref{proposition:minimal}, we start with two canonical cases of connection among three $h(h=4)\times v(v=3)$ viewports. One case is the maximum viewport connection (MVC) in the horizontal direction as illustrated in Fig. \ref{three_1}, and another is the MVC in the vertical direction as shown in Fig. \ref{three_2}.

In both cases, viewport ${l_2}$ is CV while viewports ${l_1}$ and ${l_3}$ are IVs. The number of ITs of viewport ${l_3}$ is
\begin{align}
{\mathcal{A}}({l_3})=h\cdot v-\Pi({l_2},{l_3})-\Pi({l_1},{l_2},{l_3})
\end{align}
where
\begin{align}
\Pi({l_2},{l_3})=h\cdot v-{\mathcal{A}}({l_2})-\Pi({l_1},{l_2})-\Pi({l_1},{l_2},{l_3}).
\end{align}
Then, we have
\begin{align}\label{re_1}
\Pi({l_1},{l_2})={\mathcal{A}}({l_3}).
\end{align}
With the same derivation, we have
\begin{align}\label{re_2}
\Pi({l_2},{l_3})={\mathcal{A}}({l_1}).
\end{align}
Hence, there are enough tiles $\xi({l_3})$ and $\xi({l_1})$ to respectively combine with all CTs $\xi({l_1},{l_2})$ and all CTs $\xi({l_2},{l_3})$ to form complete groups.
In the three-viewport basic cases, we can easily determine the combination relation to form complete group. Further, the combination of tiles indices is related to the size of $h$ and $v$. Analytically, in the three-viewport horizontal MVC case, for an arbitrary tile $\zeta(x,y),x\in\mathcal{H},y\in \mathcal{V}$, there exists either tile $\zeta(x+h,y)$ or tile $\zeta(x-h,y)$, and combination among them can form a complete group in definition, namely,
\begin{align}\label{ch_h}
\zeta(x,y) \cup \zeta(x+h,y) \cup \zeta(x-h,y) \overset{def}{=} {\Upsilon}.
\end{align}
When $\zeta(x+h,y)$ or $\zeta(x-h,y)$ in \eqref{ch_h} is nonexistent, we can ignore the combination with $\zeta(x+h,y)$ or $\zeta(x-h,y)$, which has no effect on the combination equation.
Likewise, in the three-viewport vertical MVC case, for an arbitrary tile $\zeta(x,y),x\in\mathcal{H},y\in \mathcal{V}$, there exists either tile $\zeta(x,y+v)$ or tile $\zeta(x,y-v)$, and
\begin{align}\label{ch_v}
\zeta(x,y) \cup \zeta(x,y+v) \cup \zeta(x,y-v) \overset{def}{=} {\Upsilon}.
\end{align}
The combination of tile indexes in \eqref{ch_h} and \eqref{ch_v} enables these three-viewport tiles to form  complete groups with number $G_{\textrm{re}}=h\cdot v$.

When viewport ${l_1}$ and ${l_3}$ have a connection with other viewports (there are more than three viewports in the graph), \eqref{re_1} and \eqref{re_2} respectively expand to
\begin{equation}
\left \{
\begin{aligned}
&\Pi({l_1},{l_2})={\mathcal{A}}({l_3})+\Pi(\sum {l_3}-{l_1}-{l_2})\\
&\Pi({l_2},{l_3})={\mathcal{A}}({l_1})+\Pi(\sum {l_1}-{l_2}-{l_3}).
\end{aligned}
\right.
\end{equation}
Note that the combination of tile indices among viewports ${l_1}, {l_2}$, and ${l_3}$ remains unchanged.

\begin{figure*}[htb]
\centering
    \includegraphics[width=0.35\textwidth]{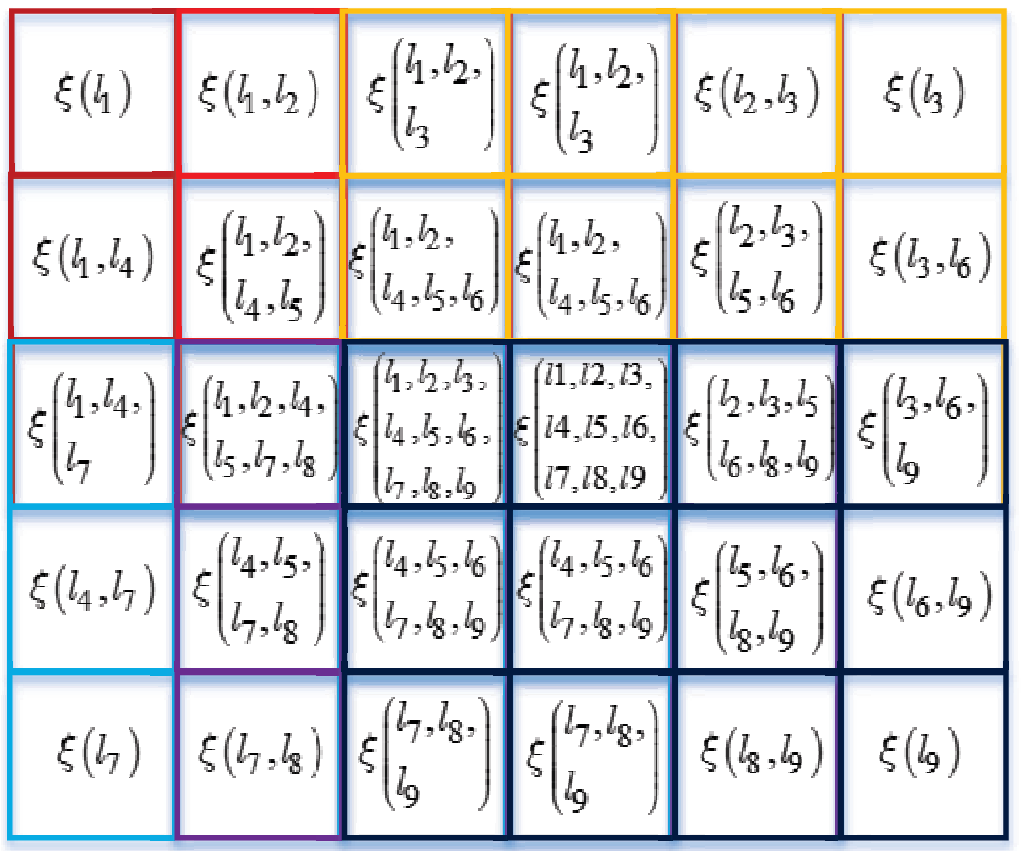}
      \caption{$6\times 5$ FMVC graph with $4\times 3$ viewport tiles, and also the FMVC graph of the example in Fig. \ref{relation}.}\label{FMVC_case}
\end{figure*}

We define the rectangular graph, which has both horizontal MVC and vertical MVC, as full MVC (FMVC) graph. For clarity, we illustrate a $6\times 5$ FMVC graph with $4\times 3$ viewport tiles in Fig. \ref{FMVC_case}.
Note that we can divide the FMVC graph into multiple overlapped MVC graphs and vertical MVC graphs, and the combination characteristics remain unchanged\footnote{Non-rectangular viewport does not have the characteristics.}. Based on the unchanged characteristics, the combination approach is suitable for a $H (H>h) \times V (V>v)$ FMVC graph. And the combination of indexes of tile $\zeta(x,y)$ is
\begin{align}\label{combination_2}
\bigcup_{j=-\left \lfloor\frac{y}{v}\right \rfloor}^{\left \lfloor\frac{V-y}{v}\right \rfloor} \bigcup_{i=-\left \lfloor\frac{x}{h}\right \rfloor}^{\left \lfloor{\frac{H-x}{h}}\right \rfloor } \zeta(x+i\cdot h,y+j\cdot v).
\end{align}
Each combined group is a complete group, and the group number is $G_{\textrm{re}}=h\cdot v$.

Note that any rectangular viewport graph has its FMVC graph according to the minimal and maximal coordinates. Hence, we easily can achieve any multi-rectangular-viewport graph through its FMVC graph, and the approach is only deleting the nonexistent viewport tiles from the FMVC graph. Further, the combination of indices of tiles remains unchanged, which is the same as \eqref{combination_2}, and the group number is also $G_{\textrm{re}}=h\cdot v$. The only difference is that $H=\max\{x\}, V=\max\{y\}$. Thus, \emph{Proposition} \ref{proposition:minimal} is proved.
\bibliographystyle{IEEEtran}
\bibliography{IEEEabrv,multicast}
\end{document}